\documentclass[10pt,journal,compsoc]{IEEEtran}

\usepackage{graphicx}
\usepackage{subfig}
\usepackage{multirow}
\usepackage{array}
\usepackage{minibox}

\usepackage{amsmath} 
\usepackage{amssymb}  
\usepackage{algorithm}
\usepackage{algpseudocode}
\usepackage{cite}
\usepackage{footnote}
\usepackage{mathtools}
\allowdisplaybreaks
\usepackage{soul}
\usepackage{flushend}
\usepackage{fmtcount}

\usepackage[colorinlistoftodos,prependcaption,textsize=tiny]{todonotes}

\begin{document}
\title{Energy-Efficient Real-Time Scheduling for Two-Type Heterogeneous Multiprocessors}
\author{Mason Thammawichai,~\IEEEmembership{Student Member,~IEEE,}
and Eric~C. Kerrigan,~\IEEEmembership{Member,~IEEE}

\IEEEcompsocitemizethanks{\IEEEcompsocthanksitem Mason Thammawichai is with the Department of Aeronautics, Imperial College London, London SW7 2AZ, UK \protect\\
E-mail: m.thammawichai12@imperial.ac.uk
\IEEEcompsocthanksitem Eric C.~Kerrigan is with the Department of Electrical \& Electronic Engineering  and the Department of Aeronautics, Imperial College London, London SW7 2AZ, UK. \protect\\
E-mail: e.kerrigan@imperial.ac.uk}}

\IEEEtitleabstractindextext{%
\begin{abstract}
We propose three novel mathematical optimization formulations that solve the same two-type heterogeneous multiprocessor scheduling problem for a real-time taskset with hard constraints. Our formulations are based on a global scheduling scheme and a fluid model. The first formulation is a mixed-integer nonlinear program, since the scheduling problem is intuitively considered as an assignment problem. However, by changing the scheduling problem to first determine a task workload partition and then to find the execution order of all tasks, the computation time can be significantly reduced. Specifically, the workload partitioning problem can be formulated as a  continuous nonlinear program for a system with continuous operating frequency, and as a continuous linear program for a practical system with a discrete speed level set. The task ordering problem can be solved by an algorithm with a complexity that is linear in the total number of tasks. The work is evaluated against existing global energy/feasibility optimal workload allocation formulations. The results illustrate that our algorithms are both feasibility optimal and energy optimal for both implicit and constrained deadline tasksets. Specifically, our algorithm can achieve up to 40\% energy saving for some simulated tasksets with constrained deadlines. The benefit of our formulation compared with existing work is that our algorithms can solve a more general class of scheduling problems due to incorporating a scheduling dynamic model in the formulations and allowing for a time-varying speed profile. Moreover, our algorithms can be applied to both online and offline scheduling schemes. 
\end{abstract}

}

\maketitle

\renewcommand{\thefootnote}{\fnsymbol{footnote}}

\renewcommand{\thefootnote}{\arabic{footnote}}
\newtheorem{theorem}{Theorem}
\newtheorem{lemma}[theorem]{Lemma}
\newtheorem{fact}[theorem]{Fact}
\newtheorem{corollary}[theorem]{Corollary}

\IEEEraisesectionheading{\section{Introduction}\label{sec:introduction}}
\IEEEPARstart{E}{fficient} energy management has become an important issue for modern computing systems due to higher computational power demands in today's computing systems, e.g.~sensor networks, satellites, multi-robot systems, as well as personal electronic devices. There are two common schemes used in modern computing energy management systems. One is dynamic power management (DPM), where certain parts of the system are turned off during the processor idle state. The other is dynamic voltage and frequency scaling (DVFS), which reduces the energy consumption by exploiting the relation between the supply voltage and power consumption. In this work, we consider the problem of scheduling real-time tasks on heterogeneous multiprocessors under a DVFS scheme with the goal of minimizing energy consumption, while ensuring that both the execution cycle requirement and timeliness constraints of real-time tasks are satisfied.

\subsection{Terminologies and Definitions}\label{sec:def}
This section provides basic terminologies and definitions used throughout the paper.

\indent\textbf{Task $T_i$}: An aperiodic task $T_i$ is defined as a triple $T_i:=(c_i,d_i,b_i)$;  $c_i$ is the required number of CPU cycles needed to complete the task, $d_i$ is the task's relative deadline and $b_i$ is the arrival time of the task. A periodic task $T_i$ is defined as a triple $T_i:=(c_i,d_i,p_i)$ where $p_i$ is the task's period. If the task's deadline is equal to its period, the task is said to have an `implicit deadline'. The task is considered to have a `constrained deadline' if its deadline is not larger than its period, i.e.\ $d_i \leq p_i$. In the case that the task's deadline can be less than, equal to, or greater than its period, it is said to have an `arbitrary deadline'. Throughout the paper, we will refer to a task as an aperiodic task model unless stated otherwise, because a periodic task can be transformed into a collection of aperiodic tasks with appropriately defined arrival times and deadlines, i.e.\ the $j^{th}$ instance of a periodic task $T_i$, where $j\geq 1$, arrives at time $(j-1)p_i$, has the required execution cycles $c_i$ and an absolute deadline at time~$(j-1)p_i+d_i$. Moreover, for a periodic taskset, we only need to find a valid schedule within its hyperperiod $\mathcal{L}$, defined as the least common multiple (LCM) of all task periods, i.e.\ the total number of job instances of a periodic task $T_i$ during the hyperperiod $\mathcal{L}$ is equal to $\mathcal{L}/p_i$. The taskset is defined as a set of all tasks. The taskset is feasible if there exists a schedule such that no task in the taskset misses the deadline.

\indent\textbf{Speed $s^r$}: The operating speed $s^r$ is defined as the ratio between the operating frequency $f^r$ of processor type-$r$ and the maximum system frequency $f_{max}$, i.e.\ $s^r:=f^r/f_{max}$, $f_{max}:=\max\left\{\max\{f^r\mid r\in R\}\right\}$, where $R:=\{1,2\}$.

\indent\textbf{Minimum Execution Time\footnote{In the literature, this is often called `worst-case execution time'. However, in the case where the speed is allowed to vary, using the term `minimum execution time' makes more sense, since the execution time increases as the speed is scaled down. For simplicity of exposition, we also assume no uncertainty, hence `worst-case' is not applicable here. Extensions to  uncertainty should be relatively straightforward, in which case $\underline{x}_i$ then becomes `minimum worst-case execution time'.} $\underline{x}_i$}: The minimum execution time $\underline{x}_i$ is the execution time of task $T_i$ when executed at the maximum system frequency $f_{max}$, i.e.\ $\underline{x}_i:=c_i/f_{max}$.

\indent\textbf{Task Density\footnote{When all tasks are assumed to have implicit deadlines, this is often called `task utilization'.} $\delta_i(s_i)$}: For a periodic task, a task density $\delta_i(s_i)$ is defined as the ratio between the task execution time and the minimum of its deadline and its period, i.e.\ $\delta_i(s_i):=c_i/(s_if_{max}\min\{d_i,p_i\})$, where $s_i$ is the task execution speed.

\indent\textbf{Taskset Density $D(s_i)$}: A taskset density $D(s_i)$ of a periodic taskset is defined as the summation of all task densities in the taskset, i.e.\ $D(s_i):=\sum_{i=1}^n \delta_i(s_i).$ The minimum taskset density $D$ is given by $D:=\sum_{i=1}^n \delta_i(1).$

\indent\textbf{System Capacity $C$}: The system capacity $C$ is defined as $C:=\sum_{r\in R}s^r_{max}m_r$, where $s^r_{max}$ is the maximum speed of processor type-$r$, i.e.\ $s^r_{max}:=f^r_{max}/f_{max},~f^r_{max}:=\max f^r$, $m_r$ is the total number of processors of type-$r$. 

\indent\textbf{Migration Scheme}: A global scheduling scheme allows task migration between processors and a partitioned scheduling scheme does not allow task migration.

\indent\textbf{Feasibility Optimal}: An algorithm is feasibility optimal if the algorithm is guaranteed to be able to construct a valid schedule such that no deadlines are missed, provided a schedule exists.

\indent\textbf{Energy Optimal}: An algorithm is energy optimal when it is guaranteed to find a schedule that minimizes the energy, while meeting the deadlines, provided such a schedule exists.

\indent\textbf{Step Function}: A function $f:X\rightarrow \mathbb{R}$ is a step (also called a piecewise constant) function, denoted $f\in\mathcal{PC}$, if there exists a finite partition $\{X_1,\ldots,X_p\}$ of $X\subseteq\mathbb{R}$ and a set of real numbers $\{\phi_1,\ldots,\phi_p\}$ such that $f(x) = \phi_i$ for all $x\in X_i$, $i\in\{1,\ldots,p\}$. 

\subsection{Related Work}
Due to the heterogeneity of the processors, one should not only consider the different operating frequency sets among processors, but also the hardware architecture of the processors, since task execution time will be different for each processor type. In other words, the system has to be captured by two aspects: the difference in operating speed sets and the execution cycles required by different tasks on different processor types.

With these aspects, fully-migration/global based scheduling algorithms, where tasks are allowed to migrate between different processor types, are not applicable in practice, since it will be difficult to identify how much computational work is executed on one processor type compared to another processor type due to differences in instruction sets, register formats, etc. Thus, most of the work related to heterogeneous multiprocessor scheduling are partition-based/non-preemptive task scheduling algorithms~\cite{yu2002,leung2004,yang2008,lee2009,jj2009,dawei2012,awan2013}, i.e.\ tasks are partitioned onto one of the processor types and a well-known uniprocessor scheduling algorithm, such as Earliest Deadline First (EDF)~\cite{Liu1973}, is used to find a valid schedule. With this scheme, the heterogeneous multiprocessor scheduling problem is reduced to a task partitioning problem, which can be formulated as an integer linear program (ILP). Examples of such work are~\cite{yu2002} and \cite{jj2009}.  

However, with the advent of ARM two-type heterogeneous multicores architecture, such as the big.LITTLE architecture~\cite{arm2013}, that supports task migrations among different core types, a global scheduling algorithm is possible. In~\cite{hoon2014,hoon2015}, the first energy-aware global scheduling framework for this special architecture is presented, where an algorithm called Hetero-Split is proposed to solve a workload assignment and a Hetero-Wrap algorithm to solve a schedule generation problem. Their framework is similar to ours, except that we adopt a fluid model to represent a scheduling dynamic, our assigned operating frequency is time-varying and the CPU idle energy consumption is also considered. 


A fluid model is the ideal schedule path of a real-time task. The remaining execution time is represented by a straight line where the slope of the line is the task execution speed. However, a practical task execution path is nonlinear, since a task may be preempted by other tasks. The execution interval of a task is represented by a line with a negative slope and a non-execution interval is represented by a line with zero slope. 

There are at least two well-known homogeneous multiprocessor scheduling algorithms that are based on a fluid scheduling model: Proportionate-fair (Pfair)~\cite{baruah96} and Largest Local Remaining Execution Time First (LLREF)~\cite{cho06}. Both Pfair and LLREF are global scheduling algorithms. By introducing the notion of \emph{fairness}, Pfair ensures that at any instant no task is one or more quanta (time intervals) away from the task's fluid path. However, the Pfair algorithm suffers from a significant run-time overhead, because tasks are split into several segments, incurring frequent algorithm invocations and task migrations. To overcome the disadvantages of quantum-based scheduling algorithms, the LLREF algorithm splits/preempts a task at two scheduling events within each time interval~\cite{cho06}. One occurs when the remaining time of an executing task is zero and it is better to select another task to run. The other event happens when the task has no laxity, i.e.\ the difference between the task deadline and the remaining execution time left is zero, hence the task needs to be selected immediately in order to finish the remaining workload in time. 

The unified theory of the deadline partitioning technique and its feasibility optimal versions, called DP-FAIR, for periodic and sporadic tasks are given in~\cite{levin2010}. Deadline Partitioning (DP)~\cite{levin2010} is the technique that partitions time into intervals bounded by two successive task deadlines, after which each task is allocated the workload and is scheduled at each time interval. A simple optimal scheduling algorithm based on DP-FAIR, called DP-WRAP, was presented in~\cite{levin2010}. The DP-WRAP algorithm partitioned time according to the DP technique and, at each time interval, the tasks are scheduled using McNaughton's wrap around algorithm~\cite{mc1959}. McNaughton's wrap around algorithm aligns all task workloads along a real number line, starting at zero, then splits tasks into chunks of length 1 and assigns each chunk to the same processor. Note that the tasks that have been split migrate between the two assigned processors. The work of~\cite{levin2010} was extended in~\cite{funk2012,fei2012} by incorporating a DVFS scheme to reduce power consumption. 

However, the algorithms that are based on the fairness notion~\cite{cho06,funaoka108,funaoka208,levin2010,funk2012,fei2012} are feasibility optimal, but have hardly been applied in a real system, since they suffer from high scheduling overheads, i.e.\ task preemptions and migrations. Recently, two feasibility optimal algorithms that are not based on the notion of fairness have been proposed. One is the RUN algorithm~\cite{regnier2011}, which uses a dualization technique to reduce the multiprocessor scheduling problem to a series of uniprocessor scheduling problems. The other is U-EDF~\cite{nelissen2012}, which generalises the earliest deadline first (EDF) algorithm to multiprocessors by reducing the problem to EDF on a uniprocessor.

Alternatively to the above methods, the multiprocessor scheduling problem can also be formulated as an optimization problem. However, since the problem is NP-hard~\cite{lawler1983}, in general, an approximated polynomial-time heuristic method is often used. An example of these approaches can be found in~\cite{chen08,xian07}, which consider energy-aware multiprocessor scheduling with probabilistic task execution times. The tasks are partitioned among the set of processors, followed with computing the running frequency based on the task execution time probabilities. Among all of the feasibility assignments, an optimal energy consumption assignment is chosen by solving a mathematical optimization problem, where the objective is to minimize some energy function. The constraints are to ensure that all tasks will meet their deadlines and only one processor is assigned to a task. In partitioned scheduling algorithms, such as~\cite{chen08,xian07}, once a task is assigned to a specific processor, the multiprocessor scheduling problem is reduced to a set of uniprocessor scheduling problems, which is well studied~\cite{chen2007}. However, a partitioned scheduling method cannot provide an optimal schedule. 

\subsection{Contribution}
The main contributions of this work are:
\begin{itemize}
\item The formulation of a real-time multiprocessor scheduling problem as an infinite-dimensional continous-time optimal control problem.
\item Three mathematical programming formulations to solve a hard real-time task scheduling problem on heterogeneous multiprocessor systems with DVFS capabilities are proposed. 
\item We provide a generalised optimal speed profile solution to a uniprocessor scheduling problem with real-time taskset. 
\item Our work is a multiprocessor scheduling algorithm that is both feasibility optimal and energy optimal. 
\item Our formulations are capable of solving a multiprocessor scheduling problem with any periodic tasksets as well as aperiodic tasksets, compared to existing work, due to the incorporation of a scheduling dynamic and a time-varying speed profile.
\item The proposed algorithms can be applied to both an online scheduling scheme, where the characteristics of the taskset is not known until the time of execution, and an offline scheduling scheme, where the taskset information is known a priori.
\item Moreover, the proposed formulations can also be extended to a multicore architecture, which only allows frequency to be changed at a cluster-level, rather than at a core-level, as explained in Section~\ref{sec:gprob}.
\end{itemize}
  
\subsection{Outline}
This paper is organized as follows: Section~\ref{sec:prob} defines our feasibility scheduling problem in detail. Details on solving the scheduling problem with finite-dimensional mathematical optimization is given in Section~\ref{sec:solve}. The optimality problem formulations are presented in Section~\ref{sec:energyOpt}. The simulation setup and results are presented in Section~\ref{sec:sim}. Finally, conclusions and future work are discussed in Section~\ref{sec:concl}.
 
\section{Feasibility Problem Formulation} \label{sec:prob}
Though our objective is to minimize the total energy consumption, we will first consider a feasiblity problem before presenting an optimality problem. 
\subsection{System model}\label{sec:sbmodel}
We consider a set of $n$ real-time tasks that are to be partitioned on a two-type heterogeneous multiprocessor system composed of $m_r$ processors of type-$r,~r\in R$. We will assume that the system supports task migration among processor types, e.g.~sharing the same instruction set and having a special interconnection for data transfer between processor types. Note that $c_i$ is the same for all processor types, since the instruction set is the same. 

\subsection{Task/Processor Assumptions} \label{sec:taskbmodel}
All tasks do not share resources, do not have any precedence constraints and are ready to start at the beginning of the execution. A task can be preempted/migrated between different processor types at any time. The cost of preemption and migration is assumed to be negligible or included in the minimum task execution times. Processors of the same type are homogeneous, i.e.\ having the same set of operating frequencies and power consumptions. Each processor's voltage/speed can be adjusted individually. 
Additionally, for an ideal system, a processor is assumed to have a continuous speed range. For a practical system, a processor is assumed to have a finite set of operating speed levels. 

\subsection{Scheduling as an Optimal Control Problem}\label{sec:gprob}
Below, we will refer to the sets $I:=\{1,\ldots,n\}$, $K^r:=\{1,\ldots,m_r\}$ and $\Gamma:=[0,L]$, where $L$ is the largest deadline of all tasks.  Note that $\forall i,\forall k,\forall r,\forall t$ are short-hand notations for $\forall i\in I,\forall k\in K^r,\forall r\in R,\forall t\in \Gamma$, respectively. The scheduling problem can therefore be formulated as the following infinite-dimensional continous-time optimal control problem:
\begin{subequations} \label{cprob}
\begin{align}
&\text{find} \quad\mathrlap{x_{i}(\cdot),a_{ik}^r(\cdot),s_{k}^r(\cdot),~\forall i\in I,k\in K^r,r\in R}\nonumber \\
	& \text{subject to } \nonumber\\
	     &\quad x_{i}(b_i) = \underline{x}_i,&&\forall i \label{he1}\\ 
			 &\quad x_{i}(t) = 0,&&\forall i,t\notin [b_i,b_i+d_i) \label{he2}\\ 
       &\quad \dot{x}_{i}(t) \geq -\sum_{r=1}^\kappa\sum_{k=1}^{m_r} a_{ik}^r(t)s_{k}^r(t),&&\forall i,t,\quad \text{a.e.} \label{he3}\\
			 &\quad \sum_{r=1}^\kappa\sum_{k=1}^{m_r} a_{ik}^r(t) \leq 1,&&\forall i,t\label{he4}\\
			 &\quad \sum_{i=1}^n a_{ik}^r(t)\leq 1,&&\forall k,r,t \label{he5}\\
			 &\quad s_k^r(t)\in S^r,&&\forall k,r,t  \label{he6}\\
			 &\quad a_{ik}^r(t) \in \{0,1\},&&\forall i,k,r,t \label{he7} \\
			 &\quad a_{ik}^r(\cdot)\in \mathcal{PC}, s_k^r(\cdot)\in \mathcal{PC},&&\forall i,k,r,t \label{he8} 
\end{align}
\end{subequations}

\noindent where the state $x_{i}(t)$ is the remaining minimum execution time of task $T_{i}$ at time $t$, the control input $s_{k}^r(t)$ is the execution speed of the $k^{th}$ processor of type-$r$ at time $t$ and the control input $a_{ik}^r(t)$ is used to indicate the processor assignment of task $T_{i}$ at time~$t$, i.e.\ $a_{ik}^r(t) = 1$ if and only if task $T_{i}$ is active on processor $k$ of type-$r$. Notice that here we formulated the problem with  speed selection at a core-level; a stricter assumption of a multicore architecture, i.e.\ a cluster-level speed assignment, is straightforward. Particularly, by replacing a core-level speed assignment $s_{k}^r$ with a cluster-level speed assignment $s^r$ in the above formulation. 

The initial conditions on the minimum execution time of all tasks and task deadline constraints are specified in~(\ref{he1}) and~(\ref{he2}), respectively. The fluid model of the scheduling dynamic is given by the differential constraint~(\ref{he3}). Constraint~(\ref{he4}) ensures that each task will be assigned to at most one non-idle processor at a time. Constraint~(\ref{he5}) quarantees that each non-idle processor will only be assigned to at most one task at a time. The speeds are constrained by~(\ref{he6}) to take on values from $S^r\subseteq[0,1]$. Constraint~(\ref{he7}) emphasis that task assignment variables are binary. Lastly,~\eqref{he8} denotes that the control inputs should be step functions. 
\begin{fact} A solution to~\eqref{cprob} where  \eqref{he3}  is satisfied with equality can be constructed from a solution to~\eqref{cprob}.
\end{fact} 
\begin{IEEEproof}
Let $(a,s,x)$ be a feasible point to~\eqref{cprob}. Let $t_i:=\min\{t\in[b_i,b_i+d_i]\mid x_i(t)\leq 0\},~\forall i$. Choose $(\tilde{a},\tilde{s},\tilde{x})$ such that (i) $\tilde{a}_{ik}^r(t)\tilde{s}_{k}^r(t) = a_{ik}^r(t)s_{k}^r(t),~\forall i,k,r,t\leq t_i$ and (ii) $\tilde{a}_{ik}^r(t)\tilde{s}_{k}^r(t) = 0,~\forall i,k,r,t > t_i$. Choose $\tilde{x}_i(0)=\underline{x}_i,~\forall i$ and $\dot{\tilde{x}}_{i}(t) = -\sum_{r=1}^\kappa\sum_{k=1}^{m_r} \tilde{a}_{ik}^r(t)\tilde{s}_{k}^r(t),~\forall i,k,r,t$. It follows that $(\tilde{a},\tilde{s},\tilde{x})$ is a solution to~\eqref{cprob} where~\eqref{he3} is an equality.
\end{IEEEproof}

\section{Solving the Scheduling Problem with Finite-dimensional Mathematical Optimization} \label{sec:solve}
The original problem~(\ref{cprob}) will be discretized by introducing piecewise constant constraints on the control inputs $s$ and $a$. Let $\mathcal{T}:=\{\tau_0, \tau_1, \ldots, \tau_{N}\}$, which we will refer to as the major grid, denote the set of discretization time steps corresponding to the distinct arrival times and deadlines of all tasks within $L$, where $0=\tau_0 < \tau_1 < \tau_2 < \cdots < \tau_{N} = L$.

\subsection{Mixed-Integer Nonlinear Program (MINLP-DVFS)}\label{sec:milp}
The above scheduling problem, subject to piecewise constant constraints on the control inputs, can be naturally formulated as an MINLP, defined below. Since the context switches due to task preemption and migration can jeopardize the performance, a variable discretization time step~\cite{gerdts2006} method is applied on a minor grid, so that the solution to our scheduling problem does not depend on the size of the discretization time step. 
Let $\{\tau_{\mu,0},\ldots,\tau_{\mu,M}\}$ denote the set of discretization time steps on a minor grid on the interval $[\tau_\mu,\tau_{\mu+1}]$ with $\tau_\mu = \tau_{\mu,0} \leq \ldots\leq\tau_{\mu,M} = \tau_{\mu+1}$, so that $\{\tau_{\mu,1},\ldots,\tau_{\mu,M-1}\}$ is to be determined for all $\mu$ from solving an appropriately-defined optimization problem.  

Let $\forall \mu$ and $\forall \nu$ be short notations for $\forall \mu\in U:=\{0,1,\ldots,N-1\}$ and $\forall \nu\in V:=\{0,1,\ldots,M-1\}$. Define the notation $[\mu,\nu]:= (\tau_{\mu,\nu}),\forall \mu,\nu$. Denote the discretized state and input sequences as
\begin{subequations}
\begin{align}
&x_{i}[\mu,\nu] := x_{i}(\tau_{\mu,\nu}),&&\forall i,\mu,\nu\\
&s_{k}^r[\mu,\nu] := s_{k}^r(\tau_{\mu,\nu}),&&\forall k,r,\mu,\nu\\
&a_{ik}^r[\mu,\nu] := a_{ik}^r(\tau_{\mu,\nu}),&&\forall i,k,r,\mu,\nu 
\end{align}
\end{subequations}

Let $s_{k}^r(\cdot)$ and $a_{ik}^r(\cdot)$ be step functions inbetween time instances on a minor grid, i.e.
\begin{subequations} \label{inputs}
\begin{align}
&s_{k}^r(t) = s_{k}^r[\mu,\nu],&&\forall t\in[\tau_{\mu,\nu},\tau_{\mu,\nu+1}),\mu,\nu \\
&a_{ik}^r(t) = a_{ik}^r[\mu,\nu],&&\forall t\in[\tau_{\mu,\nu},\tau_{\mu,\nu+1}),\mu,\nu
\end{align}
\end{subequations}

Let $\Lambda$ denote the set of all tasks within $L$, i.e.\ $\Lambda:=\{T_{i}\mid i\in I\}$. Define a task arrival time mapping $\Phi_b:\Lambda\rightarrow U$ by $\Phi_b(T_{i}) := \mu$ such that $\tau_{\mu} = b_i$ for all $T_{i}\in\Lambda$ and a task deadline mapping $\Phi_d:\Lambda\rightarrow U\cup \{N\}$ by $\Phi_d(T_{i}):=\mu$ such that $\tau_{\mu} = b_i+d_i$ for all $T_{i}\in\Lambda$. Define $\mathcal{U}_i:=\{\mu\in U\mid \Phi_b(T_i)\leq \mu<\Phi_d(T_i)\},~\forall i\in I$ and let $\forall\mu_i$ be short notation for $\forall\mu\in \mathcal{U}_i.$ 

By solving a first-order ODE with piecewise constant input, a solution of the scheduling dynamic~(\ref{he3}) has to satisfy the difference constraint
\begin{subequations}\label{dprob}
\begin{multline}
x_{i}[\mu,\nu+1]\geq x_{i}[\mu,\nu]-\\
h[\mu,\nu]\sum_{r=1}^\kappa\sum_{k=1}^{m_r}s_{k}^r[\mu,\nu]a_{ik}^r[\mu,\nu],\forall i,\mu_i,\nu. \label{eq:dyn}
\end{multline}
\noindent where $h[\mu,\nu]:=\tau_{\mu,\nu+1}-\tau_{\mu,\nu},\forall \mu,\nu.$

The discretization of the original problem~(\ref{cprob}) subject to piecewise constant constraints on the inputs~(\ref{inputs}) is therefore equivalent to the following finite-dimensional MINLP:
\begin{align}
&\text{find} \quad\mathrlap{x_{i}[\cdot],a_{ik}^r[\cdot],s_{k}^r[\cdot],h[\cdot],~\forall i\in I,k\in K^r,r\in R}\nonumber\\
	& \text{subject to (\ref{eq:dyn}) and} \nonumber\\
	     &\qquad\qquad x_{i}[\Phi_b(T_{i}),0] = \underline{x}_i,&&\forall i \label{hed1}\\ 
			 &\qquad\qquad x_{i}[\mu,\nu] = 0,&&\forall i,\mu\notin \mathcal{U}_i,\nu \label{hed2}\\ 
			 &\qquad\qquad \sum_{r=1}^\kappa\sum_{k=1}^{m_r} a_{ik}^r[\mu,\nu] \leq 1,&&\forall i,\mu,\nu\label{hed4}\\
			 &\qquad\qquad \sum_{i=1}^n a_{ik}^r[\mu,\nu]\leq 1,&&\forall k,r,\mu,\nu \label{hed5}\\
  &\qquad\qquad s_k^r[\mu,\nu]\in S^r,&&\forall k,r,\mu,\nu  \label{hed6}\\
			 &\qquad\qquad a_{ik}^r[\mu,\nu] \in \{0,1\},&&\forall i,k,r,\mu,\nu \label{hed7} \\
			 &\qquad\qquad 0 \leq h[\mu,\nu], &&\forall \mu,\nu \label{hd1}\\
			 &\qquad\qquad \sum_{\nu=0}^{M-1}h[\mu,\nu] \leq \tau_{\mu+1}-\tau_\mu, &&\forall \mu \label{hd2}
\end{align}
\end{subequations}
\noindent where~(\ref{hd1})-(\ref{hd2}) enforce upper and lower bounds on discretization time steps. 

\begin{theorem}\label{thm1} Let the size of the minor grid $M \geq \underset{r}{\max}\{m_r\}$. A solution  to~\eqref{cprob} exists if and only if a solution to~(\ref{dprob}) exists.
\end{theorem} 
\begin{IEEEproof}
Follows from the fact that if a solution exists to~\eqref{cprob}, then the Hetero-Wrap scheduling algorithm~\cite{hoon2015} can find a valid schedule with at most $m_r-1$ migrations within the cluster.~\cite[Lemma 2]{hoon2015}.

Next, we will show that $(\tilde{a}[\cdot],\tilde{s}[\cdot],\tilde{x}[\cdot],\tilde{h}[\cdot])$, a solution to~\eqref{dprob}, can be constructed from $(a(\cdot),s(\cdot),x(\cdot))$, a solution to~\eqref{cprob}. Specifically, choose $\tilde{h}[\mu,\nu]=\tau_{\mu,\nu+1}-\tau_{\mu,\nu}$ as above and $\tilde{a}_{ik}^r[\mu,\nu]$    such that  
\begin{equation}
\tilde{h}[\mu,\nu]\tilde{a}_{ik}^r[\mu,\nu] =\int_{\tau_{\mu,\nu}}^{\tau_{\mu,\nu+1}}a_{ik}^r(t)dt,~\forall i,r,\mu,\nu.
\end{equation}
Then \eqref{eq:dyn}-\eqref{hed2} are satisfied with $\tilde{x}_i[\mu,\nu]=x_i(\tau_{\mu,\nu}),~\forall i,\mu,\nu$. It follows from \eqref{he4},\eqref{he5} and~\eqref{he7} that \eqref{hed4},\eqref{hed5} and \eqref{hed7} are satified, respectively. \eqref{hed6} is satified with $\tilde{s}_k^r[\mu,\nu]=s_k^r(\tau_{\mu,\nu})$.

Suppose now we have $(\tilde{a}[\cdot],\tilde{s}[\cdot],\tilde{x}[\cdot],\tilde{h}[\cdot])$, a solution to~\eqref{dprob}. We can choose $(a(\cdot),s(\cdot),x(\cdot))$ to be a solution to~\eqref{cprob} if  the inputs are the step functions $a_{ik}^r(t) = \tilde{a}_{ik}^r[\mu,\nu]$ and $s_k^r(t) = \tilde{s}_k^r[\mu,\nu]$ when $\tilde{h}[\mu,\nu]\leq t-\tau_{\mu,\nu} < \tilde{h}[\mu,\nu+1],~\forall i,k,r,\mu,\nu$. It is simple to verify that \eqref{cprob} is satisfied by the above choice.
\end{IEEEproof}

\subsection{Computationally Tractable Multiprocessor Scheduling Algorithms}
The time to compute a solution to problem~(\ref{dprob}) is impractical even with a small problem size. However, if we relax the binary constraints in~(\ref{hed7}) so that the value of $a$ can be interpreted as the percentage of a time interval during which the task is executed (this will be denoted as $\omega$ in  later formulations), rather than the processor assignment, the problem can be reformulated as an NLP for a system with continuous operating speed and an LP for a system with discrete speed levels. The NLP and LP can be solved  at a fraction of the  time taken to solve the MINLP above. Particularly, the heterogeneous multiprocessor scheduling problem can be simplified into two steps: 
\begin{description}
\item[\textbf{STEP 1:}] \textbf{Workload Partitioning}
\item Determine the percentage of task execution times and execution speed within a time interval such that the feasibility constraints are satisfied.
\item[\textbf{STEP 2:}] \textbf{Task Ordering}
\item From the solution given in the workload partitioning step, find the execution order of all tasks within a time interval such that no task will be executed on more than one processor at a time.
\end{description}

\subsubsection{Solving the Workload Partitioning Problem as a  Continuous Nonlinear Program (NLP-DVFS)} \label{sec:nlp}

Since knowing the processor on which a task will be executed does not help in finding the task execution order, the corresponding processor assignment subscript~$k$ of the control variables $\omega$ and~$s$ is dropped to reduce the number of decision variables. Moreover, partitioning time using only a major grid (i.e. $M = 1$) is enough to guarantee a valid solution, i.e.\ the percentage of the task exection time within a major grid is equal to the sum of all percentages of task execution times in a minor grid. Since we only need a major grid, we define the notation $[\mu]:=\tau_\mu$ and $h[\mu]:=\tau_{\mu+1}-\tau_\mu$. Note that we make an assumption that $h[\mu]> 0,~\forall \mu$. We also assume that the set of allowable speed levels $S^r$ is a  closed interval given by the lower bound $s_{min}^r$ and upper bound $s_{max}^r$.

Consider now the following finite-dimensional NLP:
\begin{subequations} \label{nlpprob}
\begin{align}
&\text{find}\quad\mathrlap{x_{i}[\cdot],\omega_{i}^r[\cdot],s_{i}^r[\cdot],~\forall i\in I,r\in R}\nonumber\\
	& \text{subject to } \nonumber\\
	     &\quad\qquad x_{i}[\Phi_b(T_{i})] = \underline{x}_i,&&\forall i \label{n1}\\ 
			 &\quad\qquad x_{i}[\mu] = 0,&&\forall i,\mu\notin \mathcal{U}_i \label{n2}\\ 
       &\quad\qquad x_{i}[\mu+1]\geq x_{i}[\mu,\nu]-\nonumber\\
			 &\quad\qquad\qquad h[\mu]\sum_{r=1}^\kappa \omega_{i}^r[\mu]s_{i}^r[\mu],&&\forall i,\mu\label{n3}\\
			 &\quad\qquad \sum_{r=1}^\kappa\omega_{i}^r[\mu] \leq 1,&&\forall i,\mu\label{n5}\\
			 &\quad\qquad \sum_{i=1}^n\omega_{i}^r[\mu]\leq m_r,&&\forall r,\mu \label{n6}\\
       &\quad\qquad s_{min}^r \leq s_{i}^r[\mu] \leq s_{max}^r,&&\forall i,r,\mu \label{n7}\\
			 &\qquad\qquad 0 \leq \omega_{i}^r[\mu] \leq 1,&&\forall i,r,\mu \label{n8}
\end{align}
\end{subequations}

\noindent where $\omega^r_{i}[\mu]$ is defined as the percentage of the time interval $[\tau_{\mu},\tau_{\mu+1}]$ for which task $T_{i}$ is executing on a processor of type-$r$ at speed $s_{i}^r[\mu]$. (\ref{n5}) guarantees that a task will not run on more than one processor at a time. The constraint that the total workload at each time interval should be less than or equal to the system capacity is specified in~(\ref{n6}). Upper and lower bounds on task execution speed and  percentage of task execution time are given in~(\ref{n7}) and~(\ref{n8}), respectively. 

%
%

\subsubsection{Solving the Workload Partitioning Problem as a Linear Program (LP-DVFS)} \label{sec:lp}
 The problem~(\ref{nlpprob}) can be further simplified to an LP if the set of speed levels $S^r$ is  finite, as is often the case for practical systems. We denote with $s^r_q$ the execution speed at level $q\in Q^r:=\{1,\ldots,l_r\}$ of an $r$-type processor, where $l_r$ is the total number of speed levels of an $r$-type processor. Let $\forall q$ be short-hand  for $\forall q\in Q^r$. 

Consider now the following finite-dimensional LP:
\begin{subequations} \label{lpprob}
\begin{align}
&\text{find}\quad\mathrlap{x_{i}[\cdot],\omega_{iq}^r[\cdot],~\forall i\in I,q\in Q^r,r\in R }\nonumber\\
	& \text{subject to } \nonumber\\
	     &\qquad\qquad x_{i}[\Phi_b(T_{i})] = \underline{x}_i,&&\forall i \label{lp1}\\ 
			 &\qquad\qquad x_{i}[\mu] = 0,&&\forall i,\mu\notin \mathcal{U}_i \label{lp2}\\ 
       &\qquad\qquad x_{i}[\mu+1]\geq x_{i}[\mu]-\nonumber\\
			 &\qquad\qquad\qquad h[\mu]\sum_{r=1}^\kappa\sum_{q=1}^{l_r}\omega_{iq}^r[\mu]s^r_q,&&\forall i,\mu\label{lp3}\\
			 &\qquad\qquad \sum_{r=1}^\kappa\sum_{q=1}^{l_r} \omega_{iq}^r[\mu] \leq 1,&&\forall i,\mu\label{lp5}\\
			 &\qquad\qquad \sum_{i=1}^n\sum_{q=1}^{l_r} \omega_{iq}^r[\mu]\leq m_r,&&\forall r,\mu \label{lp6}\\
			 &\qquad\qquad 0 \leq \omega_{iq}^r[\mu] \leq 1,&&\forall i,q,r,\mu \label{lp8} 
\end{align}
\end{subequations}

\noindent where $\omega^r_{iq}[\mu]$ is the percentage of the time interval $[\tau_{\mu},\tau_{\mu+1}]$ for which task $T_{i}$ is executing on a processor of type-$r$ at a speed level $q$. Note that all constraints are similar to~(\ref{nlpprob}), but the speed levels are fixed.

\begin{theorem}\label{thm3} 
A solution to~\eqref{nlpprob} can be constructed from a solution to~\eqref{lpprob}, and vice versa, if the discrete speed set~$S^r$ is any finite subset of the closed interval $[s_{min}^r,s_{max}^r]$ with $s_{min}^r$  and $s_{max}^r$ in $S^r$ for all $r$.
\end{theorem}
\begin{IEEEproof}
Let $(\tilde{x},\tilde{\omega},\tilde{s})$ denote a solution to~\eqref{nlpprob} and $(x,\omega)$ a solution to~\eqref{lpprob}. The result follows by noting that one can  choose  $\lambda_q^r[\mu]\in [0,1]$ such that $\sum_q\lambda_q^r[\mu]s_q^r[\mu]=\tilde{s}_i^r[\mu]$, $\omega_{iq}^r[\mu]=\lambda_q^r[\mu]\tilde{\omega}_i^r[\mu]$ and $\sum_q\lambda_q^r[\mu]=1,~\forall i,q,r,\mu$ are satisfied. 
\end{IEEEproof}

\subsubsection{Task Ordering Algorithm}
This section discusses how to find a valid schedule in the task ordering step for each time interval $[\tau_{\mu},\tau_{\mu+1}]$. Since the solutions obtained in the workload partitioning step are partitioning workloads of each task on each processor type within each time interval, one might think of using McNaughton's wrap around algorithm~\cite{mc1959} to find a valid schedule for each processor within the processor type. However, McNaughton's wrap around algorithm only guarantees that a task will not be executed at the same time within the cluster. There exists a possibility that a task will be assigned to more than one processor type (cluster) at the same time.

To avoid a parallel execution on any two clusters, we can adopt the Hetero-Wrap algorithm proposed in~\cite{hoon2015} to solve a task ordering problem of a two-type heterogeneous multiprocessor platform. The algorithm takes the workload partitioning solution to STEP 1 as its inputs and returns $(\sigma_{ik}^r,\eta_{ik}^r)\in [0,1]^2,~\forall i,k,r$, which is a task-to-processor interval assignment on each cluster. Note that, for a solution to problem~(\ref{lpprob}), we define the total execution workload of a task $\omega_i^r:=\sum_q \omega_{iq}^r$ and assume that the percentage of execution times of each task at all frequency levels $\omega_{iq}^r$ will be grouped together in order to minimize the number of migrations and preemptions. In order to be self-contained, the Hetero-Wrap algorithm is given in Algorithm~\ref{HWalg}. 
\begin{algorithm}[t]
\caption{Hetero-Wrap Algorithm~\cite{hoon2015}}\label{HWalg}
\begin{algorithmic}[1]
\State \bf{INPUT:} $\omega_i^r,m_r,~\forall i,r$
\State $\sigma_{ik}^r\gets 0,~\eta_{ik}^r\gets 1,~\forall i,k,r$
\State $p_1\gets 0,~p_2\gets m_2,~k_1\gets 1,~k_2\gets m_2$
\For{$r=1,2$}
\If{$r=1$}
	\For{$i\in\{IM_a,IM_b,CP_1\}$}
		\If{$p_1=0$}
			\State $\eta_{ik_1}^r\gets w_i^r$,~$p_1\gets\eta_{ik_1}^r$
		\Else
			\If{$p_1+w_i^r\leq k_1$}
				\State $\sigma_{ik_1}^r\gets p_1-(k_1-1)$
				\State $\eta_{ik_1}^r\gets p_1+w_i^r-(k_1-1)$
				\State $p_1\gets p_1+w_i^r$
			\Else
				\State $\sigma_{ik_1}^r\gets p_1-(k-1)$ 
				\State $\eta_{i(k_1+1)}^r\gets p_1+w_i^r-k_1$
				\State $k_1\gets k_1+1$
			\EndIf
		\EndIf
	\EndFor
\Else
	\For{$i\in\{IM_a,IM_b,CP_2\}$}
		\If{$p_2=m_2$}
			\State $\sigma_{ik_2}^r\gets 1-\omega_i^r$
			\State $p_2\gets \sigma_{ik_2}^r$
		\Else
			\If{$p_2-\omega_{i}^r\geq k_2-1$}
			  \State $\sigma_{ik_2}^r\gets p_2-\omega_i^r-(k_2-1)$
				\State $\eta_{ik_2}^r\gets p_2-(k_2-1)$
				\State $p_2\gets p_2-\omega_i^r$
			\Else
				\State $\eta_{ik_2}^r\gets p_2-(k_2-1)$
				\State $k_2\gets k_2-1$
				\State $\sigma_{ik_2}\gets p_2-\omega_i^r-k_2+1$
			\EndIf
		\EndIf
	\EndFor
\EndIf

\EndFor
\State \bf{RETURN:} $(\sigma_{ik}^r,\eta_{ik}^r)\in [0,1]^2,~\forall i,k,r$
\end{algorithmic}
\end{algorithm}
Specifically, the algorithm classifies the tasks into four subsets: (i)  a set $IM_a$ of migrating tasks with $\sum_r\omega_i^r = 1$, (ii) a set $IM_b$ of migrating tasks with $\sum_r\omega_i^r<1$, (iii)  a set $CP_1$ of partitioned tasks on cluster of type-1, and (iv)  a set $CP_2$of partitioned tasks on cluster of type-2. The algorithm then employs the following simple rules:
\begin{itemize}
\item For a type-1 cluster, tasks are scheduled in the order of $IM_a,~IM_b$ and $CP_1$ using McNaughton's wrap around algorithm. That is, a slot along the number line is allocated, starting at zero, with the length equal to $m_1$ and the task is aligned with its assigned workload on empty slots of the cluster in the specified order starting from left to right. 
\item For a type-2 cluster, in the same manner, tasks are scheduled using McNaughton's wrap around algorithm, but in the order of $IM_a, IM_b$ and $CP_2$ starting from right to left. Note that the order of tasks in $IM_a$ has to be consistent with the order in a type-1 cluster.
\end{itemize}

However, the algorithm requires a feasible solution to  \eqref{nlpprob} or \eqref{lpprob}, in which $IM_b$ has at most one task, which we will call an inter-cluster migrating task. From Theorem~\ref{thm3}, we can always transform a solution to~\eqref{nlpprob} into a solution to~\eqref{lpprob}. Therefore, we only need to show that there exists a solution to~\eqref{lpprob} with at most one inter-cluster migrating tasks that lies on the vertex of the feasible region by the following facts and lemma.

\begin{fact} \label{basic} Among all the  solutions to an LP, at least one solution lies at a vertex of the feasible region. In other words, at least one solution is a basic solution.
\end{fact}  
\begin{IEEEproof}
The Fundamental Theorem of Linear Programming, which states that if a feasible solution exists, then a basic feasible solution exists~\cite[p.38]{robert2001}.
\end{IEEEproof}

\begin{fact} \label{nonbasic} A feasible solution to an LP that is not a basic solution can always be converted into a basic solution.
\end{fact}
\begin{IEEEproof}
This follows from the Fundamental Theorem of Linear Programming~\cite[p.38]{robert2001}.
\end{IEEEproof}

\begin{fact}\label{nonzero} \cite[Fact 2]{baruah2004} Consider a linear program $\min \{c^T\chi\mid A\chi\leq b, \chi\in\mathbb{R}^n\}$ for some $A\in\mathbb{R}^{(m+n)\times n}$, $b\in\mathbb{R}^{m+n}$, $c\in\mathbb{R}^{n}$. Suppose that $n$ constraints are nonnegative constraints on each variable, i.e.~$\chi_i\geq 0,~\forall i\in\{1,2,\ldots,n\}$ and the rest are $m$ linearly independent constraints. If $m<n$, then a basic solution will have at most $m$ non-zero values. 
\end{fact}
\begin{IEEEproof}
A unique basic solution can be identified by any $n+m$ linearly independent active constraints. Since there are $n$ nonnegative constraints and $m<n$, a basic solution will have at most $m$ non-zero values.  
\end{IEEEproof}

\begin{lemma} \label{intercluster} For a solution to \eqref{lpprob} that lies on the vertex of the feasible region, there will be at most one inter-cluster partitioning task.
\end{lemma}
\begin{IEEEproof}
The number of variables $\omega$ subjected to nonnegative constraint~\eqref{lp8} at each time interval of \eqref{lpprob} is $n(\sum_rl_r)$. The number of  variables $\omega$ subjected to a set of necessary and sufficient feasibility constraints~\eqref{lp5}-\eqref{lp6} is $n+2$. Note that we do not count the number of variables in \eqref{lp3} because \eqref{lp3} and \eqref{lp5} are linearly dependent constraints for a given value of $\xi_i[\mu]:=  (x_{i}[\mu]-x_{i}[\mu+1])/h[\mu]$.  If we assume that $n\geq 2$ and each processor type has at least one speed level, then it follows from Fact~\ref{nonzero} that the number of non-zero values of variable $\omega$, a solution to \eqref{lpprob} at the vertex of the feasible region, is at most $n+2$. Let $\gamma$ be the number of tasks  assigned to two processor types. Therefore, there are $2\gamma+(n-\gamma)$  entries of variable $\omega$ that are non-zero. This implies that $\gamma < 2$, i.e.\ the number of inter-cluster partitioning tasks is at most one.
\end{IEEEproof}

\begin{figure}[t]
\centering
\includegraphics[width=0.45\textwidth]{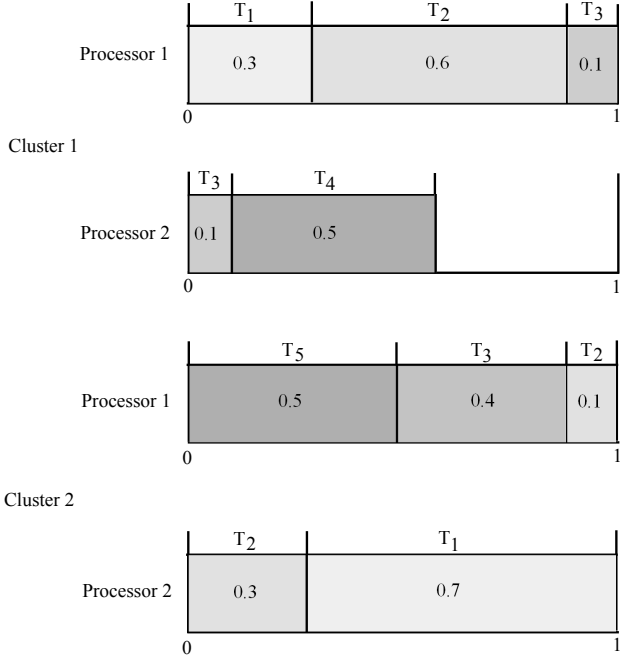}
\captionof{figure}{A feasible task schedule according to Algorithm 1.}
\label{fig:jEx}
\end{figure}

To illustrate how Algorithm~\ref{HWalg} works, consider a simple taskset in which the percentage of execution workload partition at time interval $[\tau_{\mu},\tau_{\mu+1}]$ for each task is as shown in Table~\ref{tab:ex1}. 
\begin{table}[t]
\caption{Workload partition example} \label{tab:ex1}
\centering
\begin{tabular}{|c|c|c|c|c|c|}\hline
& $T_1$ & $T_2$ & $T_3$ &$T_4$ & $T_5$\\\hline
$\omega_i^1$ & 0.3 & 0.6 & 0.2 & 0.5 & 0\\\hline
$\omega_i^2$ & 0.7 & 0.4 & 0.4 & 0 & 0.5 \\\hline
\end{tabular}
\end{table}     
A feasible schedule obtained by Algorithm~\ref{HWalg} is shown in Figure~\ref{fig:jEx}. 
For this example, $m_1 = m_2= 2$, $IM_a=\{T_1,T_2\},~IM_b=\{T_3\},~CP_1=\{T_4\}$ and $CP_2=\{T_5\}$.  

\begin{theorem}\label{thm2} If a solution  to~\eqref{cprob} exists, then a solution to~(\ref{nlpprob})/(\ref{lpprob}) exists. Furthermore, at least one valid schedule satisfying~\eqref{cprob} can be constructed from a solution to problem~(\ref{nlpprob})/(\ref{lpprob}) and the output from Algorithm~\ref{HWalg}. 
\end{theorem}
\begin{IEEEproof}
The existence of a valid schedule is proven in~\cite[Thm\ 3]{hoon2015}. It follows from Facts~\ref{basic}--\ref{nonzero} and Lemma~\ref{intercluster} that one can compute a solution with at most one inter-cluster partitioning task. Given a solution to~\eqref{nlpprob}/\eqref{lpprob} and the output from Algorithm~\ref{HWalg} for all intervals, choose $a$ to be a step function such that $a_{ik}^r(t)=1$ when $\sigma_{ik}^r[\mu,\nu]h[\mu,\nu]\leq t-\tau_{\mu,\nu}<\eta_{ik}^r[\mu,\nu]h[\mu,\nu+1]$ and $a_{ik}^r(t)=0$ otherwise, $\forall i,k,r,\mu,\nu$. Specifically, one can verify that the following condition holds
\begin{equation}
h[\mu,\nu]\omega_i^r[\mu]=\int_{\tau_{\mu,\nu}}^{\tau_{\mu,\nu+1}}{\sum_ka_{ik}^r(t)dt},~\forall i,r,\mu,\nu.
\end{equation}
Then it is straightforward to show that \eqref{cprob} is satisfied. 
\end{IEEEproof}

Note that, although, we need to solve the same multiprocessor scheduling problem with two steps in this section, the computation times to solve~(\ref{nlpprob}) or (\ref{lpprob}) is extremely fast compared to solving problem~(\ref{cprob}), i.e.\ even for a small problem, the times to compute a solution of~(\ref{dprob}) can be up to an hour, while~(\ref{nlpprob}) or (\ref{lpprob}) can be solved in milliseconds using a general-purpose desktop PC with off-the-shelf optimization solvers. Furthermore, the complexity of Algorithm~\ref{HWalg} is~$\mathcal{O}(n)$~\cite{hoon2015}.

\section{Energy Optimality} \label{sec:energyOpt}

\subsection{Energy Consumption model} \label{sec:gbmodel}
A power consumption model can be expressed as a summation of dynamic power consumption~$P_d$ and static power consumption $P_s$. Dynamic power consumption is due to the charging and discharging of CMOS gates, while static power consumption is due to subthreshold leakage current and reverse bias junction current~\cite{Rabaey:02}. The dynamic power consumption of CMOS processors at a clock frequency $f=sf_{max}$  is given by
\begin{subequations}
\begin{equation}
P_d(s) = C_{ef}V_{dd}^2sf_{max}, 
\end{equation}
where the constraint
\begin{equation}
sf_{max} \leq \zeta\frac{(V_{dd}-V_t)^2}{V_{dd}} \label{eq:s}
\end{equation}
\end{subequations}
has to be satisfied~\cite{Rabaey:02}. Here $C_{ef} > 0$ denotes the effective switch capacitance, $V_{dd}$ is the supply voltage, $V_t$ is the threshold voltage ($V_{dd}>V_t>0$\,V) and $\zeta > 0$ is a hardware-specific constant.

From~\eqref{eq:s}, it follows that if~$s$ increases, then the supply voltage $V_{dd}$ may have to increase (and if $V_{dd}$ decreases, so does $s$). In the literature, the total power consumption is often simply expressed as an increasing function of the form 
\begin{equation}
P(s):=P_d(s)+P_s=\alpha s^\beta+P_s, \label{eqn:Pd_lit}
\end{equation} where   $\alpha >0$  and $\beta \geq 1$ are hardware-dependent constants, while the static power consumption $P_s$ is assumed to be either constant or zero \cite{Li:13}.

The energy consumption of executing and completing a task $T_i$ at a constant speed $s_i$ is given by
\begin{subequations} \label{eq:ed}
\begin{align}
E(s_i) := \frac{c_i}{f_{max}} \frac{(P_d(s_i)+P_s)}{s_i} 
		 =\frac{\underline{x}_i(P_d(s_i)+P_s)}{s_i}. 
\end{align}
\end{subequations}  

In the literature, it is often assumed that $E$ is an increasing function of the operating speed. However, because  $s\mapsto 1/s$ is a decreasing function, it follows that the energy consumed might not be an increasing function if $P_s$ is non-zero; Figure~\ref{fig:ebmodel} gives an example of when the energy is non-monotonic, even if the power is an increasing function of clock frequency. 
\begin{figure}[t]
\centerline{
\subfloat[$P(s)=s^2+0.2$]{\includegraphics[scale=0.17,trim=2cm 0cm 2.5cm 0cm, clip=true]{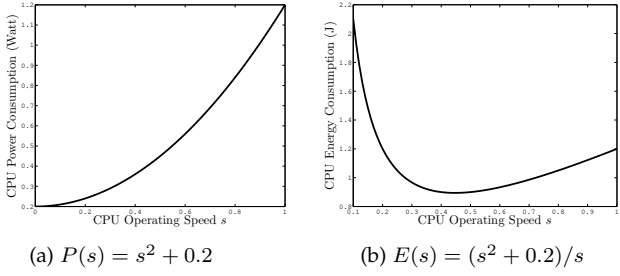}}
\subfloat[$E(s)=(s^2+0.2)/s$]{\includegraphics[scale=0.17,trim=2.5cm 0cm 0cm 0cm, clip=true]{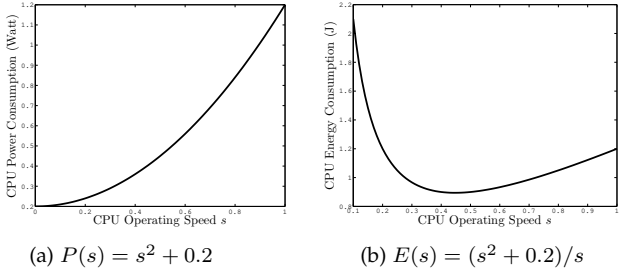}}} 
\caption{Example of a non-increasing active energy function, but where the active power consumption is an increasing function.}
\label{fig:ebmodel}
\end{figure}
This result implies the existence of a non-zero energy-efficient speed $s_{eff}$, i.e.\ the minimizer of~(\ref{eq:ed})~\cite{miyoshi02,chen06,aydin06}.  Moreover, in the work of~\cite{Zhai2004}, the non-convex relationship between the energy consumption and processor speed can be observed as a result of scaling supply voltage. 

\begin{figure}[t]
\centerline{
\subfloat[$P(s)=s^2+0.2$]{\includegraphics[scale=0.17,trim=2cm 0cm 2.5cm 0cm, clip=true]{pmodel}}
\subfloat[$E(s)=(s^2+0.2)/s$]{\includegraphics[scale=0.17,trim=2.5cm 0cm 0cm 0cm, clip=true]{emodel}}} 
\caption{Example of a non-increasing active energy function, but where the active power consumption is an increasing function.}
\label{fig:ebmodel}
\end{figure}

The total energy consumption of executing a real-time task $T_i$ can be expressed as a summation of active energy consumption and idle energy consumption, i.e.\ $E = E_{active} + E_{idle}$, where $E_{active}$ is the energy consumption when the processor is busy executing the task and $E_{idle}$ is the energy consumption when the processor is idle. The energy consumption of executing and completing a task $T_i$ at a constant speed $s_i$ is
\begin{subequations} 
\begin{align}
E(s_i) &= E_{active}(s_i) + E_{idle} \\
&=\frac{c_i}{f_{max}} \frac{(P_{active}(s_i)-P_{idle})}{s_i} + P_{idle}d_i  \\
&=\frac{\underline{x}_i(P_{active}(s_i)-P_{idle})}{s_i} + P_{idle}d_i,
\end{align}
\end{subequations} 
 
\noindent where $P_{active}(s):=P_{da}(s)+P_{sa}$ is the total power consumption in the active interval, $P_{idle}:=P_{di}+P_{si}$ is the total power consumption during the idle period. $P_{da} > 0$ and $P_{sa} \geq 0$ are dynamic and static power consumption during the active period, respectively. Similarly, $P_{di} > 0$ and $P_{si} \geq 0$ are the dynamic and static power consumption during the idle period. $P_{di}$ will be assumed to be a constant, since the processor is executing a nop (no operation) instruction at the lowest frequency $f_{min}$ during the idle interval. $P_{sa}$ and $P_{si}$ are also assumed to be constants where $P_{si} < P_{sa}$. Note that $P_{active}(s)-P_{idle}$ is strictly greater than zero.

\subsection{Optimality Problem Formulation} \label{sec:OptProb}
The scheduling problem with the objective to minimize the total energy consumption of executing the taskset on a two-type heterogeneous multiprocessor can be formulated as the following optimal control problems:
\begin{align}
&\mathrlap{\text{I) Continuous Optimal Control Problem:}}\nonumber\\
&&&\underset{\begin{subarray}{c}
          x_{i}(\cdot),a_{ik}^r(\cdot),s_{k}^r(\cdot),\\
					\forall i\in I,k\in K^r,r\in R \end{subarray}}{\text{minimize}} \quad\sum_{r,k,i}\int_0^{L} \ell^r(a_{ik}^r(t),s_{k}^r(t))dt\label{Cop}\\
&&& \text{subject to~\eqref{cprob}.} \nonumber\\
&\mathrlap{\text{II) MINLP-DVFS:}}\nonumber\\
&&& \underset{\begin{subarray}{c}
           x_{i}[\cdot],a_{ik}^r[\cdot],s_{k}^r[\cdot],h[\cdot],\\
					 \forall i\in I,k\in K^r,r\in R \end{subarray}}{\text{minimize}} \mathrlap{\quad\sum_{r,\mu,\nu,k,i}h[\mu,\nu]\ell^r(a_{ik}^r[\cdot],s_{k}^r[\cdot])}\label{Dop}\\
&&&\text{subject to~\eqref{dprob}.} \nonumber\\
&\mathrlap{\text{III) NLP-DVFS:}}\nonumber\\
&&& \underset{\begin{subarray}{c}
           x_{i}[\cdot],\omega_{i}^r[\cdot],s_{i}^r[\cdot],\\
					 \forall i\in I,r\in R \end{subarray}}{\text{minimize}} \quad\mathrlap{\sum_{r,\mu,\nu,i}h[\mu]\ell^r(\omega_{i}^r[\cdot],s_{i}^r[\cdot])} \label{NLPop}\\
&&&\text{subject to~\eqref{nlpprob}.}\nonumber\\
&\mathrlap{\text{IV) LP-DVFS}:}\nonumber\\
&&& \underset{\begin{subarray}{c}
           x_{i}[\cdot],\omega_{iq}^r[\cdot],\\
				 \forall i\in I, q\in Q^r,r\in R \end{subarray}}{\text{minimize}}\quad\mathrlap{\sum_{r,\mu,\nu,i,q}h[\mu]\ell^r(\omega_{iq}^r[\cdot],s_q^r)}\label{LPop}\\
&&&\text{subject to~\eqref{lpprob}.}\nonumber
\end{align}				
\noindent where $\ell^r(a,s):=a(P^r_{active}(s)-P^r_{idle})$. 
Note that~\eqref{LPop} is an LP, since the cost is linear in the decision variables.

\subsection{Constant or Time-varying Speed?} \label{sec:speed}
In this section, we present a result on  a general speed selection trajectory for a uniprocessor scheduling problem with a real-time taskset. With this observation about optimal speed profile, we can formulate algorithms that are able to solve a more general class of scheduling problems than in the literature.

\begin{figure}[t]
\centering
\includegraphics[width=0.47\textwidth]{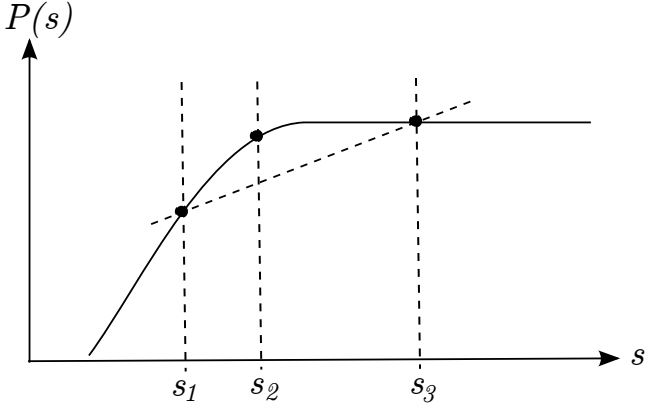}
\captionof{figure}{A time-varying speed profile is better than a constant speed profile if the power function is not convex.}
\label{fig:concave}
\end{figure}

Consider the following simple example, illustrated in Fig.~\ref{fig:concave}, where the power consumption model $P(\cdot)$ is a concave function of speed. 
Assume that $s_2$ is the lowest possible constant speed at which task $T_i$ can be finished on time, i.e.\ $\underline{x}_i=s_2d_i$. The energy consumed is $E(s_2) = P(s_2)d_i$ and the average power consumption $P(s_2)=:\bar{P}_c$. Let $\lambda\in [0,1]$ be a constant such that $s_2 = \lambda s_1 + (1-\lambda)s_3$,~$s_1 < s_2 < s_3$. Suppose $s(\cdot)$ is a time-varying speed profile such that $s(t) = s_1,~\forall t\in [0,t_1)$ and $s(t) = s_3,~\forall t\in [t_1,d_i)$. We can choose $t_1$ such that $\underline{x}_i=s_1t_1+s_3(d_i-t_1).$ The energy used in this case is $E(s_1,s_3)=t_1P(s_1)+(d_i-t_1)P(s_3)$. If we let $\lambda = t_1/d_i$, then the average power consumption $E(s_1,s_3)/d_i =: \bar{P}_{tv} = (t_1/d_i)P(s_1) + (1-(t_1/d_i))P(s_3) = \lambda P(s_1) + (1-\lambda)P(s_3).$ Since $P(\cdot)$ is concave, $P(s_2) \geq \lambda P(s_1) + (1-\lambda)P(s_3) = \bar{P}_{tv}.$ This result implies that a time-varying speed profile is better than a constant speed profile when the power consumption is concave. Notably, the result can be generalised to the case where the power model is non-convex, non-concave as well as discrete speed set.

%

\begin{theorem}\label{corol:opts} 
Let a piecewise constant speed trajectory $s^*(\cdot)$ be given that maps every time instant in a closed interval $[t_0,t_f]$ to the domain $S$ of a power function $P:S\rightarrow\mathbb{R}$. 
There exists a  piecewise constant speed trajectory $s(\cdot)$ with at most one switch such that the amount of computations done and the energy consumed is the same as using $s^*(\cdot)$, i.e.\ $s(\cdot)$ is of the form
\begin{equation}\nonumber
s(t):=
\begin{cases}
\check{s} &\forall t\in [t_0,(t_f-t_0)\lambda+t_0)\\
\hat{s} & \forall t\in [t_0+(t_f-t_0)\lambda,t_f]
\end{cases}
\end{equation}
where 
$\hat{s},\check{s}\in S$, $\lambda\in[0,1]$, such that the total amount of computations
\[
c:= \int_{t_0}^{t_f}s^*(t)dt = \int_{t_0}^{t_f}s(t)dt 
\]
and energy consumed
\[ 
E:=\int_{t_0}^{t_f}P(s^*(t))dt = \int_{t_0}^{t_f}P(s(t))dt.
\]
\end{theorem}
\begin{IEEEproof}
Let $\{\mathcal{T}_1,\ldots,\mathcal{T}_p\}$ be a partition of $[t_0,t_f]$ and  $\operatorname{range} s^*=:\{s_1,\ldots,s_p\}\subseteq S$ such that $s^*(t)=s_i$ for all $t\in\mathcal{T}_i$, $i\in\mathcal{I}:=\{1,\ldots,p\}$. Define $\Delta_i:=\int_{\mathcal{T}_i}dt$ as the size of the set $\mathcal{T}_i$ and $\lambda_i:=\Delta_i /(t_f-t_0)$, $\forall i \in \mathcal{I}$. 

It follows that $c = \sum_is_i\Delta_i$ and $E =  \sum_iP(s_i)\Delta_i$. Hence, the average speed $\bar{s}:=c/(t_f-t_0)= \sum_i\lambda_i s_i$ and  average power $\bar{P}:=E/(t_f-t_0)= \sum_i\lambda_i P(s_i)$.

Note that $\sum_i\lambda_i = 1$. This implies that $(\bar{s},\bar{P})$ is in the convex hull of the finite set $G:=\{(s_i,P(s_i))\in S\times \mathbb{R}\mid i \in \mathcal{I}\}$ with $\operatorname{vert}(\operatorname{conv}{G})\subseteq G$. Hence, there exists a $\lambda \in [0,1]$ and two points $\hat{s}$ and $\check{s}$ in $S$ with $(\hat{s},P(\hat{s}))\in \operatorname{vert}(\operatorname{conv}{G})$ and $(\check{s},P(\check{s}))\in \operatorname{vert}(\operatorname{conv}{G})$ such that $\bar{s} = \lambda\check{s}+(1-\lambda)\hat{s}$ and $\bar{P}=\lambda P(\check{s})+(1-\lambda)P(\hat{s})$. If $s$ is defined as above with these values of $\lambda$, $\hat{s}$ and $\check{s}$, then one can verify that
$
\int_{t_0}^{t_f}s(t)dt = (t_f-t_0)\bar{s}
$ 
and 
$
\int_{t_0}^{t_f}P(s(t))dt = (t_f-t_0)\bar{P}
$.
%
%
%
%
\end{IEEEproof}

The following result has already been observed in~\cite[Prop.~1]{aydin2004} and~\cite[Cor.~1]{gerards2014}.
\begin{corollary}\label{corol:optsC} Given a  processor with a convex power model and required workload within a time interval,  there exists a constant optimal speed profile if the set of speed levels~$S$ is a closed interval.
\end{corollary}
\begin{IEEEproof}
This is a special case of Theorem~\ref{corol:opts} and can be proven easily using Jensen's inequality.
\end{IEEEproof}

\begin{corollary}\label{corol:mulSOpt} An optimal speed profile  to~\eqref{Cop} can be constructed by switching between no more than two non-zero speed levels within each time interval defined by two consecutive time steps of the major grid $\mathcal{T}$.
\end{corollary}
\begin{IEEEproof}
The overall optimal speed profile can be obtained by connecting an optimal time-varying speed profile proven in Theorem~\ref{corol:opts} for each partitioned time interval. Specifically, the generalised optimal speed profile is a step function.\end{IEEEproof}

The result of the above Theorem and Corollaries can be applied directly to scheduling algorithms that adopt the DP technique such as, LLREF, DP-WRAP, as well as our algorithms  in Section~\ref{sec:solve}. Consider the problem of determining the optimal speeds at each time interval defined by two consecutive task deadlines. By subdividing time into such intervals, we can easily determine the optimal speed profile of four uniprocessor scheduling paradigms classified by power consumption and taskset models, i.e.\ (i)~a convex power consumption model with implicit deadline taskset, (ii) a convex power consumption model with constrained deadline taskset, (iii) a non-convex power consumption model with implicit deadline taskset and (iv) a non-convex power consumption model with constrained deadline taskset. Specifically, if the taskset has an implicit deadline, then the required workloads (taskset density) are equal for all time intervals; the optimal speed profiles of all schedule intervals are the same as well. Therefore, the optimal speed profile is a constant for (i) (Cor.~\ref{corol:optsC}) and a combination of two speeds for (iii) (Cor.~\ref{corol:opts}). However, for a constrained deadline taskset, the required workload varies from interval to interval, but is constant within the interval. Hence, even if the power function is (ii) convex or (iv) non-convex, the optimal speed profile is a (time-varying) piecewise constant function. In other words, for generality, a time-varying speed profile with two speed levels at each partitioned time interval is guaranteed to provide an energy optimal solution.

\begin{theorem} 
Consider the optimization problems~\eqref{Cop}--\eqref{LPop}.
An optimal speed profile for $\eqref{Cop}$ can be constructed using any of the following methods:
\begin{itemize}
\item Compute a solution to \eqref{Dop} with the lower bound on~$M$ at least twice the  bound in Theorem~\ref{thm1}.
\item If the active power function $P_{active}$ is convex and the speed level sets are closed intervals, compute a solution to \eqref{NLPop}.  If there is  more than one inter-cluster partitioning task, then  the (finite) range of the optimal speed profile should be used to define and compute a solution to \eqref{LPop} with at most  one inter-cluster partitioning task. This process is concluded with Algorithm~\ref{HWalg}.
\item If the speed level sets are finite, compute a solution to \eqref{LPop} with at most  one inter-cluster partitioning task, followed with Algorithm~\ref{HWalg}.
\end{itemize}
\end{theorem}
\begin{IEEEproof}
Follows from the choices of selecting $a$ and $s$ as in the proofs of Theorem~\ref{thm1} and Theorem~\ref{thm2}. The cost of all problems are then equal.
\end{IEEEproof}

\section{Simulation Results} \label{sec:sim}

\subsection{System, Processor and Task models} \label{sec:spm}
\begin{figure*}[t]
\centerline{
\subfloat[big core Processor]{\includegraphics[width=0.5\textwidth]{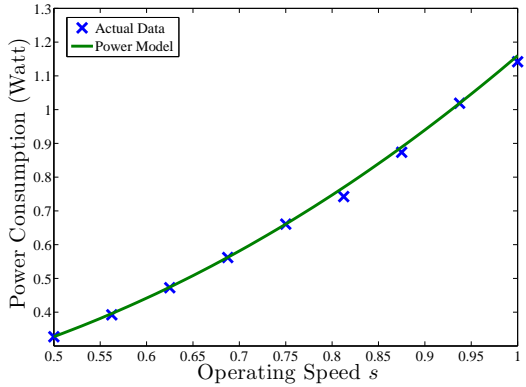}}
\subfloat[LITTLE core Processor]{\includegraphics[width=0.5\textwidth]{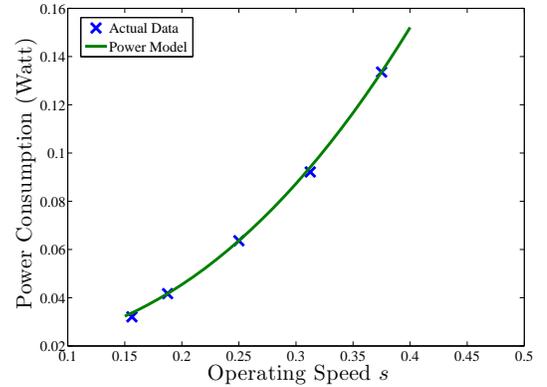}}}
\caption{Actual Data versus Fitted model}
\label{fig:fitplot}
\end{figure*}

The energy efficiency of solving the above optimization problems is evaluated on the ARM big.LITTLE architecture, where a big core provides faster execution times, but consumes more energy than a LITTLE core. The details of the ARM Cortex-A15 (big) and Cortex-A7 (LITTLE) core, which have been validated in~\cite{hoon2014}, are given in Tables~\ref{tab:pbmodel1} and~\ref{tab:pbmodel2}. 
\begin{table*}[!t]
\small
\caption{ARM Cortex-A15 (big) Processor Details~\cite{hoon2014}} \label{tab:pbmodel1}
\centering
\begin{tabular}{|*{10}{c|}}\hline
Voltage (V) & 0.93 & 0.96 & 1.0 & 1.04 & 1.08 & 1.1 & 1.15 & 1.2 & 1.23\\ \hline
Freq. (MHz) & 800 & 900 & 1000 & 1100 & 1200 & 1300 & 1400 & 1500 & 1600  \\ \hline
Speed  & 0.5 & 0.5625 & 0.625 & 0.6875 & 0.75 & 0.8125 & 0.875 & 0.9375 & 1.0  \\ \hline
Power (mW) & 327 & 392 & 472 & 562 & 661 & 742 & 874 & 1,019 & 1,142 \\ \hline
\end{tabular}
\end{table*}
\begin{table*}[!t]
\small
\caption{ARM Cortex-A7 (LITTLE) Processor Details~\cite{hoon2014}} \label{tab:pbmodel2}
\centering
\begin{tabular}{|*{6}{c|}}\hline
Voltage (V) & 0.9 & 0.94 & 1.01 & 1.09 & 1.2\\ \hline
Freq. (MHz) & 250 & 300 & 400 & 500 & 600 \\ \hline
Speed  & 0.1563 & 0.1875 & 0.25 & 0.3125 & 0.375 \\ \hline
Power (mW) & 32 & 42 & 64 & 92 & 134\\ \hline
\end{tabular}
\end{table*}
The active power consumption models, obtained by a polynomial curve fitting to the generic form~(\ref{eqn:Pd_lit}), are shown in Table~\ref{tab:fitbmodel}. 
\begin{table*}[t]
\caption{ARM Processor Power Consumption models} \label{tab:fitbmodel}
\centering
\addtocounter{footnote}{1}
\begin{tabular}{|c|c|c|}\hline
Processor & Active Power model & MAPE$^{\decimal{footnote}}$ \\\hline
big & $P_{active}(s):=1063.9s^{2.2}+95.9075$ & 0.9283  \\\hline
LITTLE &$P_{active}(s):=1103.17s^{2.3034}+18.3549$ & 1.4131 \\\hline
\end{tabular}
\end{table*}
The plots of the actual data versus the fitted models are shown in Fig.~\ref{fig:fitplot}. 
The idle power consumption was not reported, thus we will assume this to be a constant strictly less than the lowest active power consumption, namely $P_{idle}=70$~mW for the big core and $P_{idle}=12$~mW for the LITTLE core. To illustrate that our formulations are able to solve a broader class of multiprocessor scheduling problems than others optimal algorithms reported in the literature, we consider periodic taskset models with both implicit and constrained deadlines. However, a more general taskset model such as an arbitary deadline taskset, where the deadline could be greater than the period, a sporadic taskset model, where the inter-arrival time of successive tasks is at least~$p_i$ time units, and an aperiodic taskset can be solved by our algorithms as well. To guarantee the existence of a valid schedule, the minimum taskset density has to be less than or equal to the system capacity. Moreover, a periodic task needs to be able to be executed on any processor type. Specifically, the minimum task density should be less than or equal to the lowest capacity of all processor types, i.e. $\delta_i(1)\leq 0.375$ for this particular architecture.

\subsection{Comparison between Algorithms}
For a system with a continuous speed range, four algorithms are compared: (i) MINLP-DVFS, (ii) NLP-DVFS, (iii) GWA-SVFS, which represents a global energy/feasibility-optimal workload allocation with constant frequency scaling scheme at a core-level and (iv) GWA-NoDVFS, which is a global scheduling approach without frequency scaling scheme. For a system with discrete speed levels, four algorithms are compared: (i) LP-DVFS, (ii) GWA-NoDVFS, (iii) GWA-DDiscrete and (iv) GWA-SDiscrete, which represent global energy/feasibility-optimal workload allocation with time-varying and constant discrete frequency scaling schemes, respectively. Note that GWA-SVFS, GWA-NoDVFS, GWA-DDiscrete and GWA-SDiscrete are based on the mathematical optimization formulation proposed in~\cite{hoon2014}, but adapted to our framework, for which details are given below.

GWA-SVFS/GWA-NoDVFS: Given $m_r$ processors of type-$r$ and $n$ periodic tasks, determine a constant operating speed for each processor $s^r_k$ and the workload ratio $y_{ik}^r$ for all tasks within hyperperiod $\mathcal{L}$ that solves:
\begin{subequations} \label{GSprob}
\begin{align}
& \underset{\begin{subarray}{c}
           s^r_k,y_{ik}^r,\\
					 i\in I,k\in K^r,r\in R \end{subarray}}{\text{minimize}}\quad\mathrlap{\sum_{r,i,k} \mathcal{L}_i\ell^r(\delta_{ik}^r(s^r_k),s^r_k)} \\
	& \text{subject to } \nonumber\\
	     &\qquad \sum_{r=1}^\kappa\sum_{k=1}^{m_r} y_{ik}^r = 1,&&\forall i \label{gs1}\\
			 &\qquad \sum_{r=1}^\kappa\sum_{k=1}^{m_r} \delta_{ik}^r(s^r_k) \leq 1,&&\forall i \label{gs2}\\
			 &\qquad \sum_{i=1}^n\delta_{ik}^r(s^r_k) \leq 1,&&\forall k,r \label{gs3}\\
			 &\qquad 0 \leq y_{ik}^r \leq 1,&&\forall i,k,r \label{gs4}\\
			 &\qquad s^r_{min} \leq s^r_k \leq s^r_{max},&&\forall k,r&&\text{(GWA-SVFS)} \label{gs5}\\
			 &\qquad s^r_k = s^r_{max},&&\forall k,r&&\text{(GWA-NoDVFS)}\label{gs6} 
\end{align}
\end{subequations}

\noindent where $y_{ik}^r$ is the ratio of the workload of task $T_i$ on processor $k$ of type-$r$, $\delta_{ik}^r(s^r_k)$ is the task density on processor $k$ type-$r$ defined as $\delta_{ik}^r(s^r_k):=y_{ik}^rc_{i}/(s^r_kf_{max}\min\{d_i,p_i\}$) and $\mathcal{L}_i:=\mathcal{L}\min\{d_i,p_i\}/p_i$. Note that when $d_i=p_i$ as in the case of an implicit deadline taskset $\mathcal{L}_i = \mathcal{L},~\forall i$. (\ref{gs1}) ensures that all tasks will be allocated the amount of required execution time. The constraint that a task will not be executed on more than one processor at the same time is specified in~(\ref{gs2}). (\ref{gs3}) asserts that the assigned workload will not exceed processor type capacity. Upper and lower bounds on the workload ratio of a task are given in~(\ref{gs4}). The difference between GWA-SVFS and GWA-NoDVFS lies in restricting a core-level operating speed $s^r_k$ to be either a continuous variable~(\ref{gs5}) or fixed at the maximum value~(\ref{gs6}). 

GWA-DDiscrete: Given $m_r$ processors of type-$r$ and $n$ periodic tasks, determine a percentage of the task workload $y_{iq}^r$ at a specific speed level for all tasks within hyperperiod $\mathcal{L}$ that solves:
\begin{subequations} \label{Ddiscrete}
\begin{align}
& \underset{\begin{subarray}{c}
           y_{iq}^r\\
					 i\in I,q\in Q^r,r\in R \end{subarray}}{\text{minimize}}\quad\mathrlap{\sum_{r,i,q} \mathcal{L}_i\ell^r(\delta_{iq}^r(y_{iq}^r),s^r_q)}\\
	& \text{subject to } \nonumber\\
	     &\qquad\qquad \sum_{r=1}^\kappa\sum_{q=1}^{l_r} y_{iq}^r = 1,&&\forall i \label{dd1}\\
			 &\qquad\qquad \sum_{r=1}^\kappa\sum_{q=1}^{l_r} \delta_{iq}^r(y_{iq}^r) \leq 1,&&\forall i \label{dd2}\\
			 &\qquad\qquad \sum_{i=1}^n\sum_{q=1}^{l_r} \delta_{iq}^r(y_{iq}^r) \leq m_r,&&\forall r \label{dd3}\\
			 &\qquad\qquad 0 \leq y_{iq}^r \leq 1,&&\forall i,q,r \label{dd4}
\end{align}
\end{subequations}

\noindent where $y_{iq}^r$ is the percentage of workload of task $T_i$ on processor type-$r$ at speed level~$q$, $\delta_{iq}^r(y_{iq}^r)$ is the task density on processor type-$r$ at speed level $q$, i.e.\ $\delta_{iq}^r(y_{iq}^r):=y_{iq}^rc_i/(s^r_qf_{max}\min\{d_i,p_i\}$). Constraint~(\ref{dd1}) guarantees that the total execution workload of a task is allocated. Constraint~(\ref{dd2}) assures that a task will be executed only on one processor at a time. Constraint~(\ref{dd3}) ensures that each processor type workload capacity is not violated. Constraint~(\ref{dd4}) provides upper and lower bounds on a percentage of task workload at specific speed level.

\footnotetext[\value{footnote}]{Mean Absolute Percentage Error (MAPE)\\
MAPE$:=\frac{1}{k}\sum_{i=1}^k\frac{|F(z(i))-y(i)|}{|y(i)|}\times 100$, where $\frac{|F(z(i))-y(i)|}{|y(i)|}$ is the magnitude of the relative error in the $i^{th}$ measurement, $z\mapsto F(z)$ is the estimated function, $z$ is the input data, $y$ is the actual data and $k$ is the total number of fitted points.}

GWA-SDiscrete: Given $m_r$ processors of type-$r$ and $n$ periodic tasks, determine a percentage of task workload $y_{iq}^r$ at a specific speed level and a processor speed level selection $z_q^r$ for all tasks within hyperperiod $\mathcal{L}$ that solves:
\begin{subequations} \label{Sdiscrete}
\begin{align}
& \underset{\begin{subarray}{c}
           y_{ikq}^r,z_{kq}^r\\
					 i\in I,k\in K^r,q\in Q^r,r\in R \end{subarray}}{\text{minimize}}\quad\mathrlap{\sum_{r,i,q} \mathcal{L}_i\ell^r(\delta_{ikq}^r(y_{ikq}^r)z_{kq}^r,s^r_q)}\\
	& \text{subject to } \nonumber\\
	     &\qquad\qquad \sum_{r=1}^\kappa\sum_{q=1}^{l_r}\sum_{k=1}^{m_r} y_{ikq}^rz_{kq}^r = 1,&&\forall i \label{sd1}\\
			 &\qquad\qquad \sum_{q=1}^{l_r} z_{kq}^r = 1,&&\forall k,r \label{sd2} \\
			 &\qquad\qquad \sum_{r=1}^\kappa\sum_{q=1}^{l_r} \delta_{ikq}^r(y_{ikq}^r)z_{kq}^r \leq 1,&&\forall i \label{sd3}\\
			 &\qquad\qquad \sum_{i=1}^n\sum_{q=1}^{l_r} \delta_{ikq}^r(y_{ikq}^r)z_{kq}^r \leq 1,&&\forall k,r \label{sd4}\\
			 &\qquad\qquad 0 \leq y_{ikq}^r \leq 1,&&\forall i,k,q,r \label{sd5} \\
			 &\qquad\qquad z_{kq}^r \in \{0,1\},&&\forall k,q,r \label{sd6}
\end{align}
\end{subequations}

\noindent where $y_{ikq}^r$ is the workload partition of task $T_i$ of processor $k$ of an $r$-type at speed level~$q$, $z_{kq}^r$ is a speed level selection variable for processor $k$ of an $r$-type , i.e.\ $z_{kq}^r = 1$ if a speed level $q$ of an $r$-type processor is selected and $z_{kq}^r = 0$ otherwise. Constraint~(\ref{sd1}),~(\ref{sd3})--(\ref{sd5}) are the same as the GWA-DDiscrete. Constraint~(\ref{sd2}) assures that only one speed level is selected. Constraint~(\ref{sd6}) emphasises that the speed level selection variable is a binary.

Note that GWA-SVFS and GWA-NoDVFS are NLPs, GWA-DDiscrete is an LP and GWA-SDiscrete is an MINLP. Moreover, the formulation of GWA-DDiscrete allows a processor to run with a time-varying combination of constant discrete speed levels, while GWA-SVFS and GWA-SDiscrete only allow a constant execution speed for each processor.

\begin{table*}[!ht]
\caption{Implicit deadline tasksets for simulation} \label{tab:Imptaskset}
\centering
\begin{tabular}{|c|l||c|l|}\hline
$D$& Taskset &$D$ &Taskset\\ \hline
0.50 & (1,5),(1,10),(4,20) & 2.00 & (1,5),(3,10),(7,20),(7,20),(7,20),(7,20),(2,20) \\\hline
0.75 & (1,5),(1,10),(5,20),(4,20) & 2.25 & (1,5),(3.5,10),(3.5,10),(6,20),(7,20),(7,20),(7,20)\\\hline
1.00 & (1,5),(1,10),(7,20),(7,20) & 2.50 & (1,5),(3,10),(3,10),(3,10),(7,20),(7,20),(7,20),(7,20)\\\hline
1.25 & (1,5),(1,10),(6,20),(6,20),(7,20) & 2.75 & (1,5),(3.5,10),(3.5,10),(3,10),(7,20),(7,20),(7,20),(7,20),(3,20) \\\hline
1.50 & (1,5),(3,10),(6,20),(7,20),(7,20) & 3.00 & (1,5),(3,10),(3,10),(3,10),(3,10),(7,20),(7,20),(7,20),(7,20),(4,20) \\\hline
1.75 & (1,5),(2,10),(6,20),(7,20),(7,20),(7,20) & 3.25 & (1,5),(3.5,10),(3.5,10),(3.5,10),(3.5,10),(7,20),(7,20),(7,20),(7,20),(5,20)\\\hline
3.50 & \multicolumn{3}{l|}{(1,5),(3.5,10),(3.5,10),(3,10),(3,10),(7,20),(7,20),(7,20),(7,20),(7,20),(5,20)} \\\hline
3.75 & \multicolumn{3}{l|}{(1,5),(3.5,10),(3.5,10),(3,10),(3,10),(7.5,20),(7.5,20),(7.5,20),(7.5,20),(7.5,20),(7.5,20) }\\\hline
4.00 & \multicolumn{3}{l|}{(1,5),(3.5,10),(3.5,10),(3,10),(3,10),(7.5,20),(7.5,20),(7.5,20),(7.5,20),(7.5,20),(7.5,20),(5,20)}\\\hline
4.25 & \multicolumn{3}{l|}{(1,5),(3.75,10),(3.75,10),(3.75,10),(3.75,10),(7.5,20),(7.5,20),(7.5,20),(7.5,20),(7.5,20),(7.5,20),(6,20)}\\\hline
\multicolumn{4}{l}{Note: (i) The first parameter of a task is $\underline{x}_i$; $c_i$ can be obtained by multiplying $\underline{x}_i$ by $f_{max}$.}\\
\multicolumn{4}{l}{\qquad~~(ii) Since the task period is the same as the deadline, the last parameter of the task model is dropped.} 
\end{tabular}
\end{table*}

\subsection{Simulation Setup and Results}
For simplicity and without loss of generality, consider the case where independent real-time tasks are to be executed on two-type processor architectures, for which the details are given in Section~\ref{sec:spm}. The MINLP formulations were modelled using ZIMPL~\cite{Koch04} and solved with SCIP~\cite{Achterberg09}. The LP and NLP formulations were solved with SoPlex~\cite{Wunderling1996} and Ipopt~\cite{wachter2006}, respectively. The value of the minor grid discretization step $M$ is chosen according to Theorem~\ref{thm1}.  

For implict deadline tasksets, we consider the system composed of two big cores and six LITTLE cores, which has a system capacity of 4.25=(2+2.25). The total energy consumption of each taskset with a minimum taskset density varying from 0.5 to system capacity with a step of~0.25, given in Table~\ref{tab:Imptaskset}, are evaluated. 

\begin{figure*}[t]
\centerline{\subfloat[A system with continuous speed range]{\includegraphics[width=0.5\textwidth]{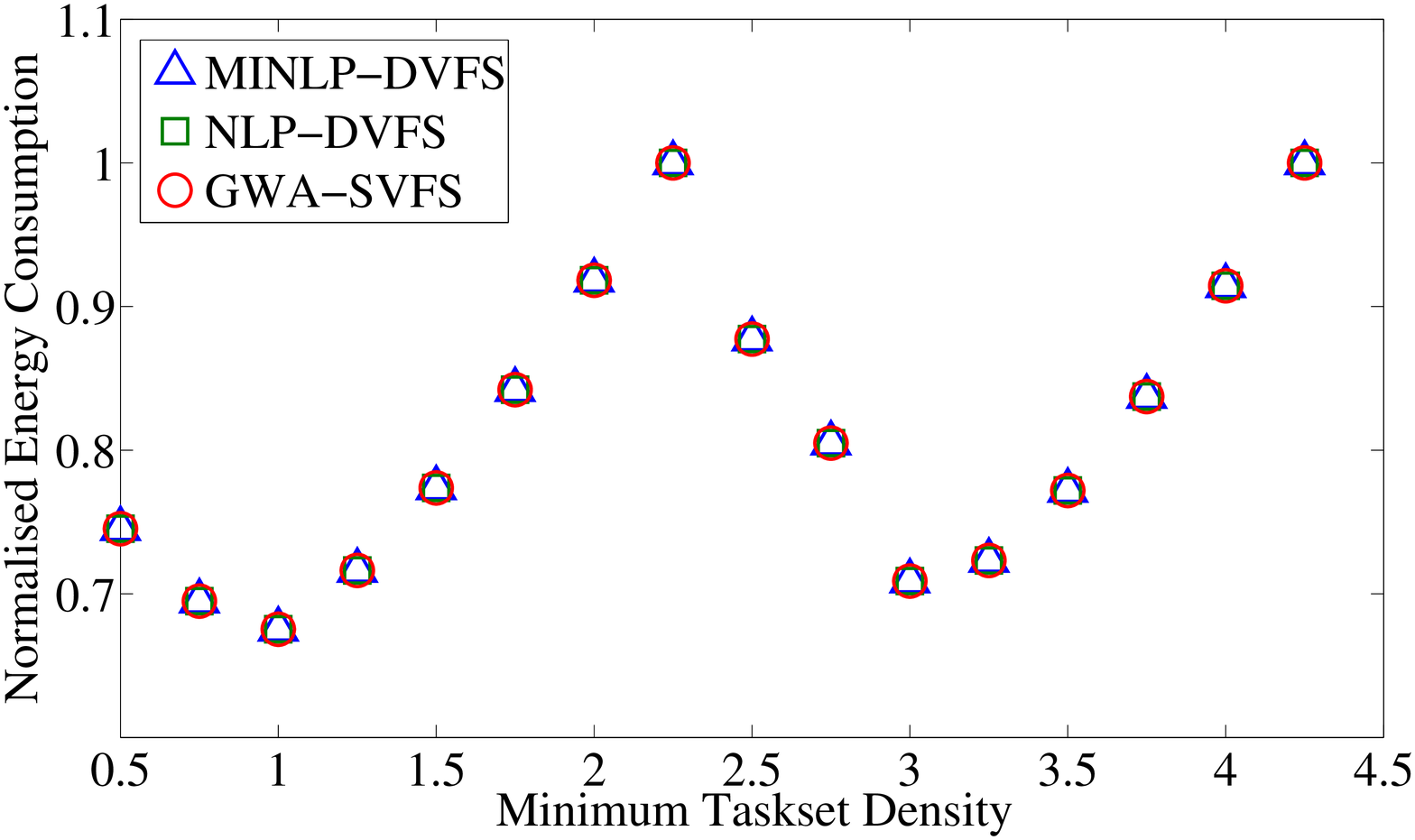}}
\subfloat[A system with discrete speed levels]{\includegraphics[width=0.5\textwidth]{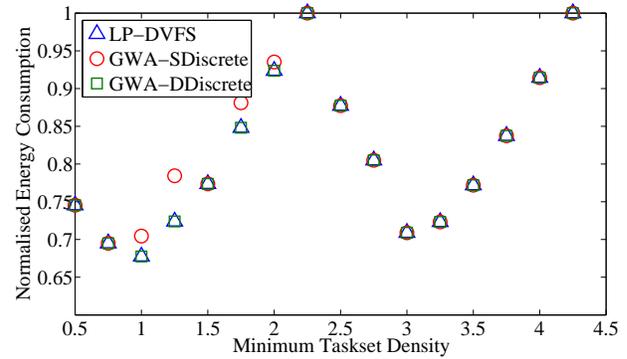}}}
\caption{Simulation results for scheduling real-time tasks with implicit deadlines}
\label{fig:ResultPlot}
\end{figure*}

Figure~\ref{fig:ResultPlot}a shows simulation results for scheduling a real-time taskset with implicit deadlines on an ideal system. 
The minimum taskset density $D$ is represented on the horizontal axis. The vertical axis is the total energy consumption normalised by GWA-NoDVFS, where less than~1 means the algorithm does better than GWA-NoDVFS. 

The three algorithms with a DVFS scheme, i.e.\ MINLP-DVFS, NLP-DVFS, and GWA-SVFS, produce the same optimal energy consumption, though both of our algorithms allow the operating speed to vary with time compared with a constant frequency scaling scheme, used by GWA-SVFS. The simulation results suggest that the optimal speed is a constant, rather than time-varying, for an implicit deadline taskset that has a constant workload over time. This result complies with Corollary~\ref{corol:optsC}. Moreover, the little core, which only has 37.5\% computing power compared with the big core and consumes considerably less power even when running at full speed, will be selected by the optimizer before considering the big cores. This is why we can see two upwards parabolic curves in the figures, where the first one corresponds to the case where only little cores in the system are selected, while both core types are selected in the second, which happens when the minimum taskset density is larger than the little-core cluster's capacity. 

However, for a practical system, where a processor has discrete speed levels, the constant speed assignment is not an optimal strategy. As can be observed in Figure~\ref{fig:ResultPlot}b, the LP-DVFS and GWA-DDiscrete are energy optimal, while the GWA-SDiscrete is not. The results imply that to obtain an energy optimal schedule, a time-varying combination of discrete speed levels is necessary. 

\begin{figure*}[t]
\centerline{\subfloat[A system with continous speed range]{\includegraphics[width=0.5\textwidth]{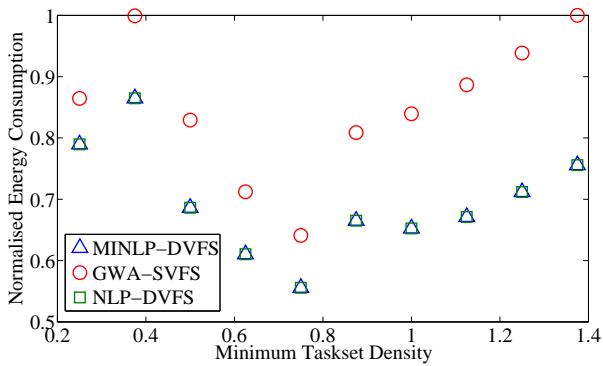}}
\subfloat[A system with discrete speed levels]{\includegraphics[width=0.5\textwidth]{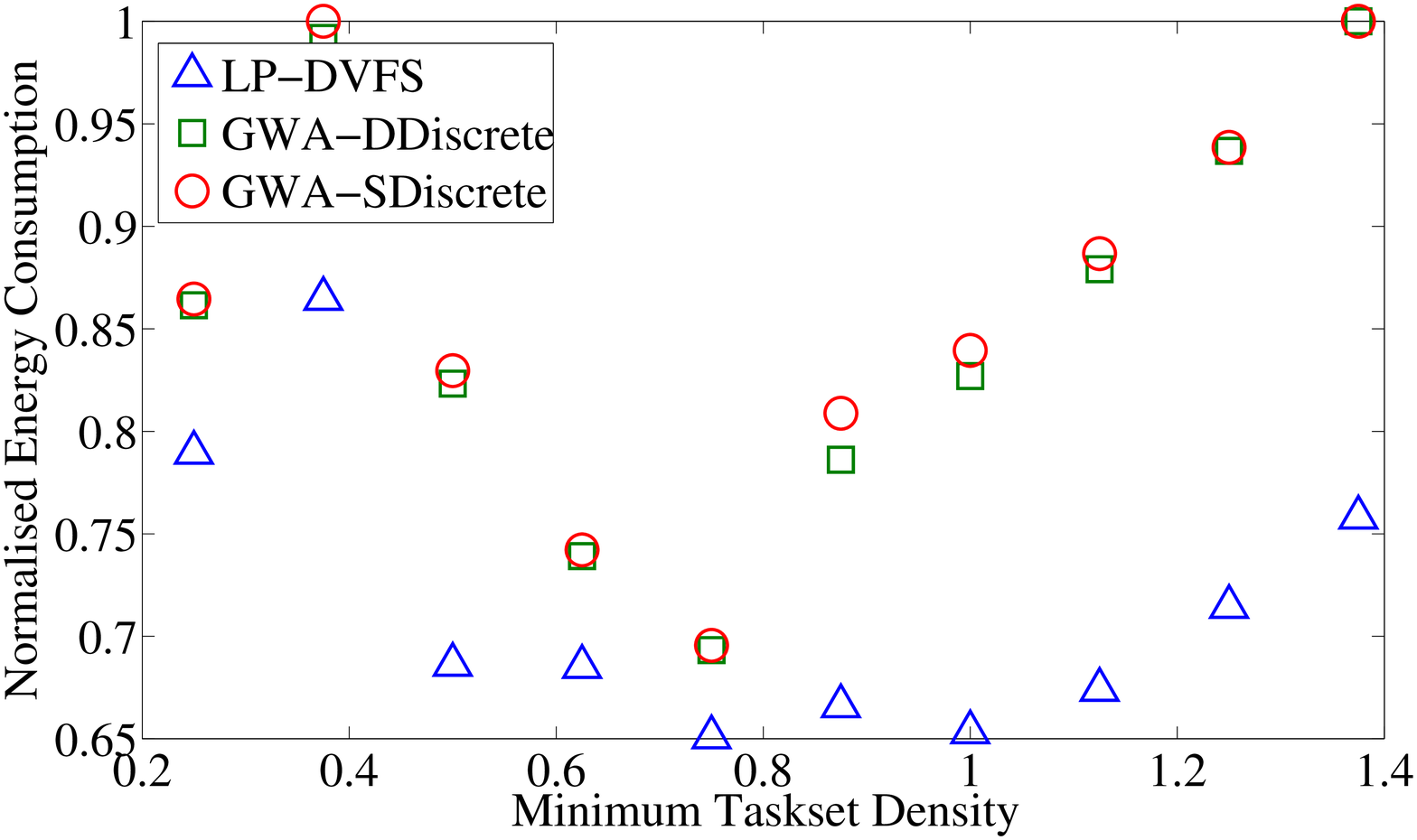}}}
\caption{Simulation results for scheduling real-time tasks with constrained deadlines}
\label{fig:ResultPlot2}
\end{figure*}

For a real-time taskset with constrained deadlines, we consider a system with one big core and one LITTLE core, i.e.\ a system capacity of 1.375=(1+0.375). The simulation results of executing each taskset, listed in Table~\ref{tab:ctaskset}, are shown in Figures~\ref{fig:ResultPlot2}, where the total energy consumption normalised by GWA-NoDVFS is on the vertical axis. 
\begin{table}[t]
\caption{Constrained deadline tasksets for simulation} \label{tab:ctaskset}
\centering
\begin{tabular}{|c|l|}\hline
$D$& Taskset\\ \hline
0.250 & (0.9375,5,10),(0.625,10,10) \\\hline
0.375 & (1.5625,5,10),(0.625,10,10) \\\hline
0.500 & (1.875,5,10),(1.25,10,10)\\\hline
0.625 & (1.875,5,10),(1,5,10),(0.5,10,10) \\\hline
0.750 & (1.875,5,10),(1.625,5,10),(0.5,10,10) \\\hline
0.875 & (1.875,5,40),(1.75,5,40),(6,40,40) \\\hline
1.000 & (1.875,5,40),(1.875,5,40),(0.5,5,40),(6,40,40) \\\hline
1.125 & (1.875,5,40),(1.5,5,40),(1.3125,5,40),(6,40,40),(1.5,40,40) \\ \hline
1.25 & (1.875,5,40),(1.875,5,40),(1.5625,5,40),(6,40,40),(1.5,40,40)\\ \hline
1.375 & (1.875,5,40),(1.875,5,40),(1.875,5,40),(6,40,40),(4,40,40)\\ \hline
\end{tabular}
\end{table}
It can be seen from the plots that for a taskset with a piecewise constant and time-varying workload, i.e.\ constrained deadlines, GWA-SVFS, GWA-DDiscrete and GWA-SDiscrete cannot provide an optimal energy consumption, while our algorithms are optimal. This is because time is incorporated in our formulations, which provides benefits for solving a scheduling problem with a time-varying workload as well as a constant workload. 

Lastly, it has to be mentioned that the energy saving percentage varies with the taskset, which implies that the number on the plots shown here can be varied, but the significant outcomes stay the same.

\section{Conclusions}  \label{sec:concl}
This work presents multiprocessor scheduling as an optimal control problem with the objective of minimizing the total energy consumption. We have shown that the scheduling problem is computationally tractable by first solving a workload partitioning problem, then a task ordering problem. The simulation results illustrate that our algorithms are both feasibility optimal and energy optimal when compared to an existing global energy/feasibility optimal workload allocation algorithm. Moreover, we have shown via proof and simulation that a constant frequency scaling scheme is enough to guarantee optimal energy consumption for an ideal system with a constant workload and convex power function, while this is not true in the case of a time-varying workload or a non-convex power function. For a practical system with discrete speed levels, a time-varying speed assignment is necessary to obtain an optimal energy consumption in general. 

For future work, one could incorporate a DPM scheme and formulate the problem as a multi-objective optimization problem to further reduce energy consumption of a system. Extending the idea presented here to cope with uncertainty in a task's execution time using feedback is also possible. Though our work has been focused on minimizing the energy consumption, the framework could be easily applied to other objectives such as leakage-aware, thermal-aware and communication-aware scheduling problems. Numerically efficient methods could also be developed to solve optimization problems defined here. 

\bibliographystyle{IEEEtran}        
\bibliography{HMS}

\begin{IEEEbiography}[{\includegraphics[width=1in,clip,keepaspectratio]{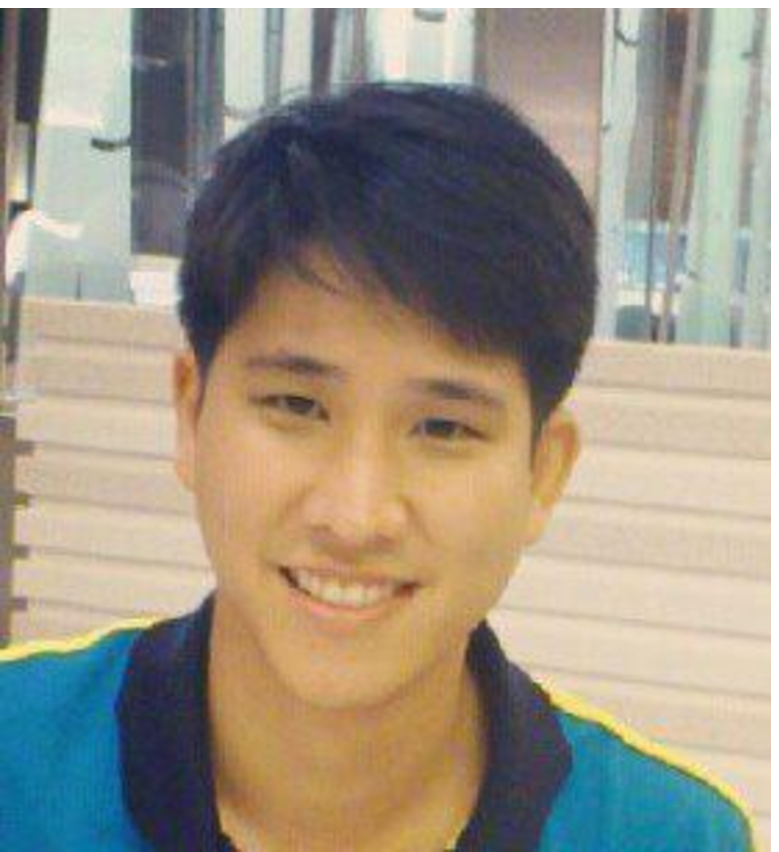}}]{Mason Thammawichai}
received the BS degree in Computer Engineering from University of Wisconsin-Madison, USA and the MSc in Avionic Systems from University of Sheffield, UK. He is currently a PhD student at Imperial College London, UK. His main areas of research are real-time scheduling, mathematical optimization, optimal control and intelligent multi-agent systems.
\end{IEEEbiography}

\begin{IEEEbiography}[{\includegraphics[width=1in,clip,keepaspectratio]{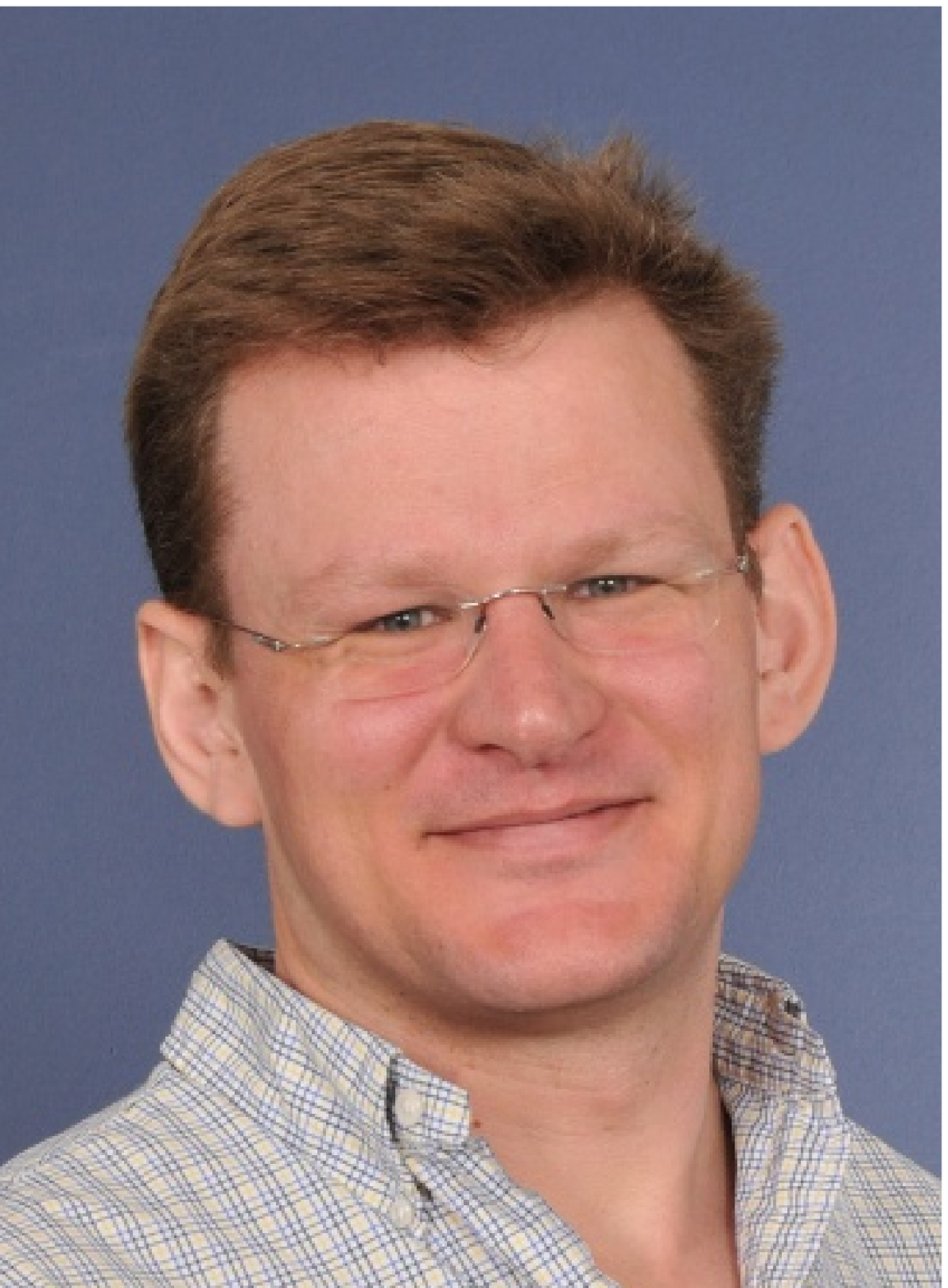}}]{Eric C.\ Kerrigan}
(S'94-M'02) received a PhD from the University of Cambridge in 2001 and has been a faculty member at Imperial College London since 2006. His research is on efficient numerical methods and  computing architectures for solving advanced optimization, control and estimation problems arising  in  aerospace, renewable energy and computing systems. He is on the IEEE Control Systems Society Conference Editorial Board and is an associate editor of the IEEE Transactions on Control Systems Technology and Control Engineering Practice.
\end{IEEEbiography}

\end{document}